\documentclass[twocolumn,english,amsmath,amssymb,superscriptaddress,nofootinbib,twocolumn]{revtex4-1}
\usepackage{mathptmx}
\usepackage[T1]{fontenc}
\usepackage[latin9]{inputenc}
\usepackage{amsmath}
\usepackage{amssymb}
\usepackage{graphicx}
\usepackage{babel}
\usepackage[normalem]{ulem}

\usepackage{color}

\begin{document}

\title{Active Frequency Measurement on Superradiant Strontium Clock Transitions}

\author{Yuan Zhang}
\email{yzhuaudipc@zzu.edu.cn}
\address{Henan Key Laboratory of Diamond Optoelectronic Materials and Devices, Key Laboratory of Material Physics Ministry of Education, School of Physics and Microelectronics, Zhengzhou University, Daxue Road 75, Zhengzhou 450052 China} 

\author{Chongxin Shan}
\email{cxshan@zzu.edu.cn}
\address{Henan Key Laboratory of Diamond Optoelectronic Materials and Devices, Key Laboratory of Material Physics Ministry of Education, School of Physics and Microelectronics, Zhengzhou University, Daxue Road 75, Zhengzhou 450052 China} 

\author{Klaus M{\o}lmer}
\email{moelmer@phys.au.dk}

\address{Center for Complex Quantum Systems, Department of Physics and Astronomy, Aarhus University, Ny Munkegade 120, DK-8000 Aarhus C, Denmark}

\begin{abstract}
We develop a stochastic mean-field theory to describe active frequency measurements of pulsed superradiant emission, studied in recent experiments with strontium-87 atoms trapped in an optical lattice inside an optical cavity [M. Norcia, et al., Phys. Rev. X \textbf{8}, 21036 (2018)]. Our theory reveals the intriguing dynamics of atomic ensembles with multiple transition frequencies, and it reproduces the superradiant beats signal, noisy power spectra, and frequency uncertainty in remarkable agreement with the experiments.  Moreover, by reducing the number of atoms, elongating the superradiant pulses and shortening the experimental duty cycle, we predict a short-term frequency uncertainty $9\times10^{-16} \sqrt{\tau/s}$, which makes active frequency measurements with superradiant transitions comparable with the record performance of current frequency standards [M. Schioppo, et. al., Nat. Photonics, \textbf{11}, 48 (2017)].  Our theory combines cavity-quantum electrodynamics and quantum measurement theory, and it can be readily applied to explore conditional quantum dynamics and describe frequency measurements for other processes such as  steady-state superradiance and superradiant Raman lasing.
\end{abstract}
\maketitle

\paragraph{Introduction}
Optical clocks possess superior precision and accuracy compared to
their counterparts in the microwave domain \citep{ADLudlow}.
The pursuit of optical clocks follows two paths with either single trapped ions, which allow long interrogation times and achieve impressive frequency resolution \citep{WHOdksy}, or ensembles of neutral alkaline-earth atoms, which can provide a better signal-to-noise ratio and which can be trapped in optical lattices to reduce the effect of Doppler broadening \citep{MMB}. Beyond quantum metrology \citep{JYe}, trapped ions and atoms are useful for the exploration of quantum computing \citep{JJGarcia} and simulation, as well as  many-body spin physics \citep{MJMartin,KKim}. 

Most optical clocks are based on a passive scheme \citep{ADLudlow},
where atoms are excited by a driving field and its frequency is matched to the atomic transition frequency by monitoring the atomic population dynamics, e.g., via fluorescence detection.  Alternatively, in an active scheme, the atomic transition frequency is measured by comparing the signal emitted by the atoms with a reference laser. While the passive scheme has superior long-term stability and absolute accuracy, the active scheme offers a wide detection bandwidth and dynamical range. The active scheme has been implemented in the microwave domain using, e.g., hydrogen masers \citep{MAWeiss}. Recently, M. Norcia, et al. \citep{MANorcia} demonstrated active frequency measurements of the  pulsed superradiant radiation emitted by strontium atoms on the optical clock transition in the optical domain. 

\begin{figure}
\begin{centering}
\includegraphics[scale=0.5]{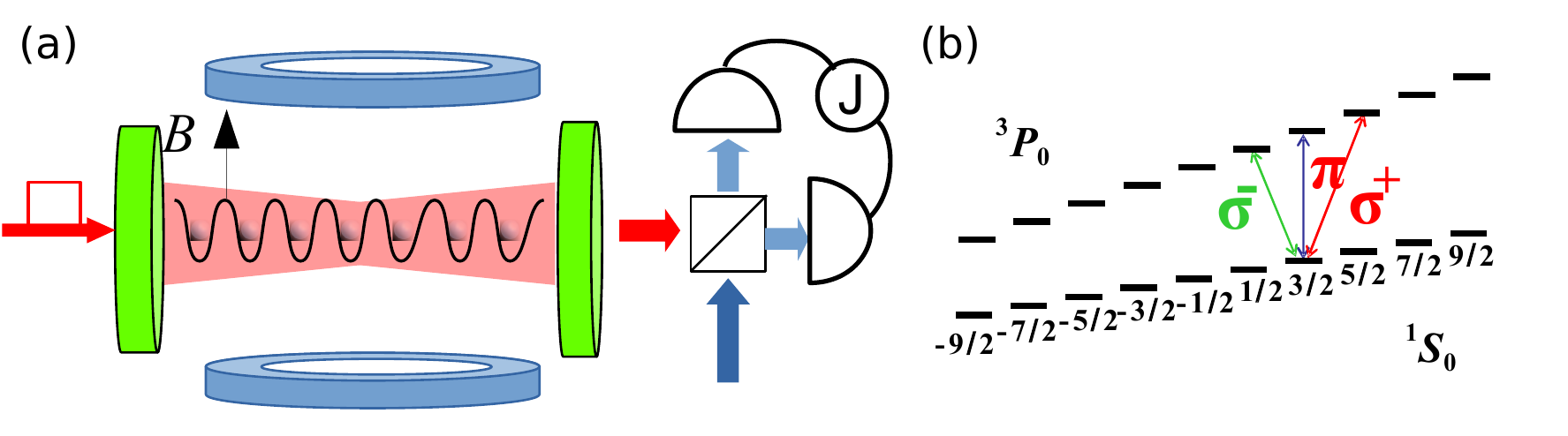}
\par\end{centering}
\caption{\label{fig:system-Zeeman} Panel (a) shows an ensemble of  ${}^{87}{\rm Sr}$ lattice trapped atoms in a cavity, which are subject to a magnetic field and balanced heterodyne detection by photon-counting on the beams mixing the superradiance from the cavity (horizontal fat red arrow) and a reference laser  (vertical long blue arrow).  Panel (b) shows hyper-fine levels labeled by $m_{F}=-F,-F+1,...F$ ($F=9/2$) of the electronic ground $^{1}S_{0}$ and excited $^{3}P_{0}$ states and three kinds of transitions $\hat{\sigma}^{+},\pi,\sigma^{-}$ ( with $\Delta m_{F}=+1,0,1$). By defining the quantization axis with a vertical magnetic field, the cavity mode with vertical polarization couples only to the atomic $\pi$-transitions. }
\end{figure}

The system studied in \citep{MANorcia} consists of more than $10^{5}$ strontium-87 atoms trapped in a one-dimensional optical lattice inside an optical cavity, see Fig. \ref{fig:system-Zeeman}(a). The ${}^{87}{\rm Sr}$ atoms have nuclear spin $F=9/2$, and  subject to a magnetic field, the hyper-fine levels of the electronic ground $^{1}S_{0}$ and excited state $^{3}P_{0}$ are Zeeman split and give rise to a number of $\sigma^{\pm}$ and $\pi$ transitions, see Fig. \ref{fig:system-Zeeman}(b).  The atoms couple to a cavity mode with vertical polarization via the $\pi$ transitions, which can be labeled with the quantum number $m_{F}=-F,-F+1,...F$. The average frequency of transitions labelled by  $m_F$ and $-m_F$ is independent of the magnetic field and can thus be used as an absolute frequency reference, while the frequency difference of the adjacent transitions is proportional to the magnetic field and can thus be used for magnetometry.

The procedure adopted in the experiment \citep{MANorcia}  is shown in Fig. \ref{fig:system-Zeeman}(a) and can be summarized as follows: i) a laser pulse drives the cavity to excite the atoms; ii) the excited atoms interact collectively with the cavity mode resulting in pulsed superradiance; iii) in the heterodyne detection, the superradiance signal interferes with a reference laser and is measured with photo-detectors; iv) the Fourier-transform of the photo-current leads to the power spectral intensity. The peaks in the spectrum reflect the beat-notes in the interference signal, from which the atomic transition frequencies can be inferred. Following this procedure, a frequency uncertainty of $8\times 10^{-16}$ is  achieved after one second of integration time \citep{MANorcia}.We note that in \citep{MANorcia-2}, the initial atomic excitation is realized by chirped adiabatic passage.

In this article, we combine cavity-quantum-electrodynamics (QED) and quantum measurement theory to accurately model the physics involved in the above high-precision frequency measurement and to determine its ultimate frequency uncertainty. Our theory combines the conditioned dynamics of quantum emitters subject to continuous measurements \citep{HWWiseman} with the accurate theory of light-matter interaction, superradiance \citep{DMeiser,YZhang-2} and strong coupling \citep{HMabuchi}. Our simulations show that the frequency uncertainty can be reduced by one to two orders of magnitude by using longer superradiance pulses and by reducing the time needed for single measurements. The optimized uncertainty becomes comparable with the current record  $6\times 10^{-17}$ at one second \citep{MSchioppo}.

\paragraph{Stochastic Master Equation for Conditional Dynamics}
In the following, we present the stochastic master equation for the conditioned density operator $\hat{\rho}$ and we derive a stochastic mean field theory to effectively describe the system of many emitters. 
Our starting point is the evolution due to different physical mechanisms, 
\begin{equation}
\frac{\partial}{\partial t}\hat{\rho}=\left(\frac{\partial}{\partial t}\hat{\rho}\right)_{p}+\left(\frac{\partial}{\partial t}\hat{\rho}\right)_{s}+\left(\frac{\partial}{\partial t}\hat{\rho}\right)_{d}.\label{eq:stochastic-master-equation}
\end{equation}
The first term,
\begin{equation}
\left(\frac{\partial}{\partial t}\hat{\rho}\right)_{p}=-\frac{i}{\hbar}\left[\hat{H}_{c}+\hat{H}_{d},\hat{\rho}\right]+\kappa\mathcal{D}\left[\hat{a}\right]\hat{\rho},\label{eq:preparation}
\end{equation}
specifies the cavity mode Hamiltonian $\hat{H}_{c}=\hbar\omega_{c}\hat{a}^{+}\hat{a}$ with frequency $\omega_{c}$, photon creation $\hat{a}^{+}$ and annihilation operator $\hat{a}$ , and the driving of the cavity mode, $\hat{H}_{d}=\sqrt{\kappa_{1}}\hbar\Omega\left(t\right)e^{i\omega_{d}(t)t}\hat{a}+\mathrm{h.c.}$,
by a laser pulse with a frequency $\omega_{d}$ and  a time-dependent strength $\Omega\left(t\right)$. The Lindblad damping term describes the photon loss with a rate $\kappa=\kappa_{1}+\kappa_{2}$ due to the left ($\kappa_1$) and right ($\kappa_2$) mirror, and the superoperator is defined as $\mathcal{D}\left[\hat{o}\right]\hat{\rho}=\hat{o}\hat{\rho} \hat{o}^{+}-\left\{ \hat{o}^{+}\hat{o},\hat{\rho}\right\} /2$
(for any operator $\hat{o}$). Here, we assume that the laser pulse enters the cavity through the left mirror and  the loss rates through the mirrors are equal $\kappa_1=\kappa_2=2\pi\times72.5$ kHz. 

The second part of Eq. (\ref{eq:stochastic-master-equation}),
\begin{align}
 & \left(\frac{\partial}{\partial t}\hat{\rho}\right)_{s}=-\frac{i}{\hbar}\left[\hat{H}_{a}+\hat{H}_{a-c},\hat{\rho}\right]+\sum_{i=1}^{10} \gamma_{i}\sum_{k=1}^{N_{i}}\mathcal{\mathcal{D}}\left[\hat{\sigma}_{i,k}^{-}\right]\hat{\rho}.\label{eq:interaction}
\end{align}
specifies the Hamiltonian $\hat{H}_{a}=\hbar\sum_{i=1}^{10}(\omega_{i}/2)\sum_{k=1}^{N_{i}}\hat{\sigma}_{i,k}^{z}$ of ten sub-ensembles of $N_{i}$ atoms, which are associated with the ten $\pi$-transitions with frequencies
$\omega_{i=m_{F}}=\omega_{a}+\Delta_{B}m_{F}$ and  Pauli operator $\hat{\sigma}_{i,k}^{z}$.  The frequencies $\omega_{i=m_{F}}$ are given by the intrinsic atomic transition frequency $\omega_a/2\pi=429.5$ THz (corresponding to a wavelength of $698$ nm) and the constant  \textbf{$\Delta_{B}=108.4\times B$} Hz for the static magnetic field  $B$ in Gauss \citep{MANorcia}.
The Hamiltonian $\hat{H}_{a-c}=\hbar\sum_{i}g_{i}\left(\hat{a}^{+}\sum_{k}\hat{\sigma}_{i,k}^{-}+\sum_{k}\hat{\sigma}_{i,k}^{+}\hat{a} \right)$ describes the atom-cavity mode coupling with the lowering   $\hat{\sigma}_{i,k}^{-}$ and raising $\hat{\sigma}_{i,k}^{+}$ operators and the strengths $g_{i=m_{F}}=g_{0}m_{F}/\sqrt{F\left(F+1\right)}$, which are determined by a constant $g_{0}=2\pi\times2.41$ Hz  and Clebsch-Gordan coefficients. The Lindblad term describes the excited-state decay with the rate $\gamma_{i}$, which is assumed to be the same for all the excited states, $\gamma_i = \gamma_0=2\pi$ mHz in our simulations. Here, we have ignored the negligible dephasing rate in the optical lattice clock system.

The third part of Eq. (\ref{eq:stochastic-master-equation}) describes the measurement back-action due to the heterodyne detection. Unlike the experiment  \citep{MANorcia}, here, we consider balanced heterodyne detection and measure the field amplitude of two beams obtained by even mixing with a strong reference laser. The difference of the photon-currents in the two detectors is 
\begin{align}
J\left(t\right) & =\sqrt{\eta\kappa_2}2\mathrm{Re}\left[e^{-i\Delta t}\left\langle \hat{a}\right\rangle \right]+ dW/dt \label{current},
\end{align}
which is proportional to the cavity field amplitude $\left\langle \hat{a}\right\rangle$
but is dominated by the detector shot-noise contribution $dW/dt$. In every step of the simulations, we generate a random number $dW$ following a normal distribution with a variance $dW\left(t\right)^{2}=dt$ and
a mean $E\left[dW\left(t\right)\right]=0$, and we employ
\begin{equation}
\left(\frac{\partial}{\partial t}\hat{\rho}\right)_{d}=\frac{dW}{dt}\sqrt{\eta\kappa_2}\bigl[e^{-i\Delta t}\left(\hat{a}-\left\langle \hat{a}\right\rangle \right)\hat{\rho}+\hat{\rho} e^{i\Delta t}\left(\hat{a}^{+}-\left\langle \hat{a}\right\rangle ^{*}\right)\bigr]\label{eq:measurment-backaction}
\end{equation}
to account for the measurement backaction on the system \citep{HWWiseman}. In Eq. (\ref{eq:measurment-backaction}), $\Delta=\omega_{l}-\omega_{c}$ denotes the difference of the reference laser frequency $\omega_{l}$ and the cavity mode frequency $\omega_{c}$.

To simulate the dynamics of the tens of thousands of atoms studied in the experiments and to account properly for the collective atom-cavity mode interaction, we solve the stochastic master equation (\ref{eq:stochastic-master-equation}) with second-order mean-field theory \cite{YZhang-2,YZhang-3}, equivalent to the cluster expansion approach \citep{HAMLeymann}. In this theory, the master equation (\ref{eq:stochastic-master-equation}) is applied to derive the equation of evolution, $\partial_{t}\left\langle \hat{o}\right\rangle =\mathrm{tr}\left\{ \hat{o}\partial_{t}\hat{\rho}\right\} $, for the expectation value $\left\langle \hat{o}\right\rangle$ of any system observable $\hat{o}$. Following this procedure, we obtain the equation of evolution for the expectation value of the intra-cavity photon number:
\begin{align}
 & \frac{\partial}{\partial t}\left\langle \hat{a}^{+}\hat{a}\right\rangle =-2\mathrm{Im\sqrt{\kappa_{1}}}\Omega\left(t\right)e^{i\omega_{d}t}\left\langle \hat{a}\right\rangle \nonumber \\
 &  -\kappa\left\langle \hat{a}^{+}\hat{a}\right\rangle -2\sum_{i}g_{i}N_{i}\mathrm{Im}\left\langle \hat{a}\hat{\sigma}_{i}^{+}\right\rangle \nonumber \\
 & +\frac{dW}{dt}\sqrt{\eta\kappa_2} 2\mathrm{Re}\left[e^{i\Delta t}\left(\left\langle \hat{a}^{+}\hat{a}^{+}\hat{a}\right\rangle -\left\langle \hat{a}\right\rangle ^{*}\left\langle \hat{a}^{+}\hat{a}\right\rangle \right)\right], \label{eq:photon-number}
\end{align}
which depends on the expectation value of the cavity field amplitude $\left\langle \hat{a}\right\rangle $ through the driving (the first line) and the atom-photon correlation $\left\langle \hat{a}\hat{\sigma}_{i}^{+}\right\rangle $ through the atom-cavity mode interaction  (the second line), and also on higher order correlations $\left\langle \hat{a}^{+}\hat{a}^{+}\hat{a}\right\rangle $ through the heterodyne detection (the third line). We assume identical correlations between all $N$ atoms and the cavity field, and to close the equations, we approximate the third-order correlations $\left\langle \hat{o}\hat{p}\hat{q}\right\rangle $ with
lower-order correlations $\left\langle \hat{o}\right\rangle \left\langle \hat{p} \hat{q}\right\rangle +\left\langle \hat{p}\right\rangle \left\langle \hat{o}\hat{q}\right\rangle +\left\langle \hat{q}\right\rangle \left\langle \hat{o}\hat{p}\right\rangle -2\left\langle \hat{o}\right\rangle \left\langle \hat{p}\right\rangle \left\langle \hat{q}\right\rangle $
($\hat{o},\hat{p},\hat{q}$ for any operator) \citep{DMeiser}. The expectation value of the cavity field annihilation operator then obeys the equation
\begin{align}
 & \frac{\partial}{\partial t}\left\langle \hat{a}\right\rangle =-i\tilde{\omega}_{c}\left\langle \hat{a}\right\rangle -i\sqrt{\kappa_{1}}\Omega^{*}\left(t\right)e^{-i\omega_{d}t}-i\sum_{i}g_{i}N_{i}\left\langle \hat{\sigma}_{i}^{-}\right\rangle \nonumber \\
 & +\frac{dW}{dt}\sqrt{\eta\kappa_{2}}\left[e^{i\Delta t}\left(\left\langle \hat{a}^{+}\hat{a}\right\rangle -\left|\left\langle \hat{a}\right\rangle \right|^{2}\right)+e^{-i\Delta t}\left(\left\langle \hat{a}^{2}\right\rangle -\left\langle \hat{a}\right\rangle ^{2}\right)\right]\label{eq:amplitude}
\end{align}
with  the complex frequency $\tilde{\omega}_{c}=\omega_{c}-i \kappa/2$. The equations for $\left\langle \hat{a}\hat{\sigma}_{i}^{+}\right\rangle $ and other terms, e.g., the atomic population difference $\left\langle \hat{\sigma}_{i}^{z}\right\rangle $ and polarization $\left\langle \hat{\sigma}_{i}^{-}\right\rangle $, are
detailed in the Appendix \ref{sec:observables}.

\paragraph{Frequency Measurement using Superradiance Pulses}

\begin{figure}
\begin{centering}
\includegraphics[scale=0.45]{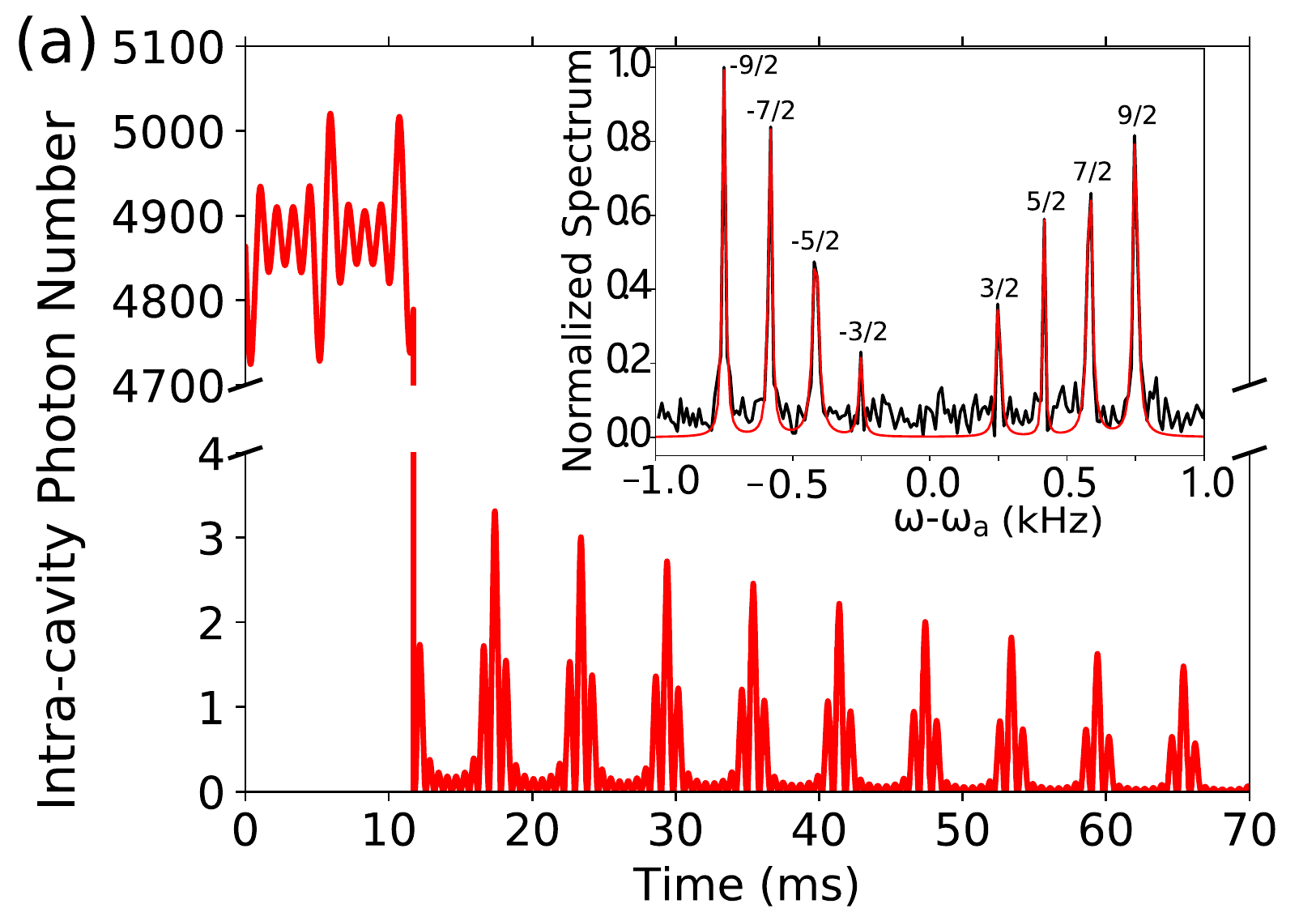}
\par\end{centering}
\begin{centering}
\includegraphics[scale=0.43]{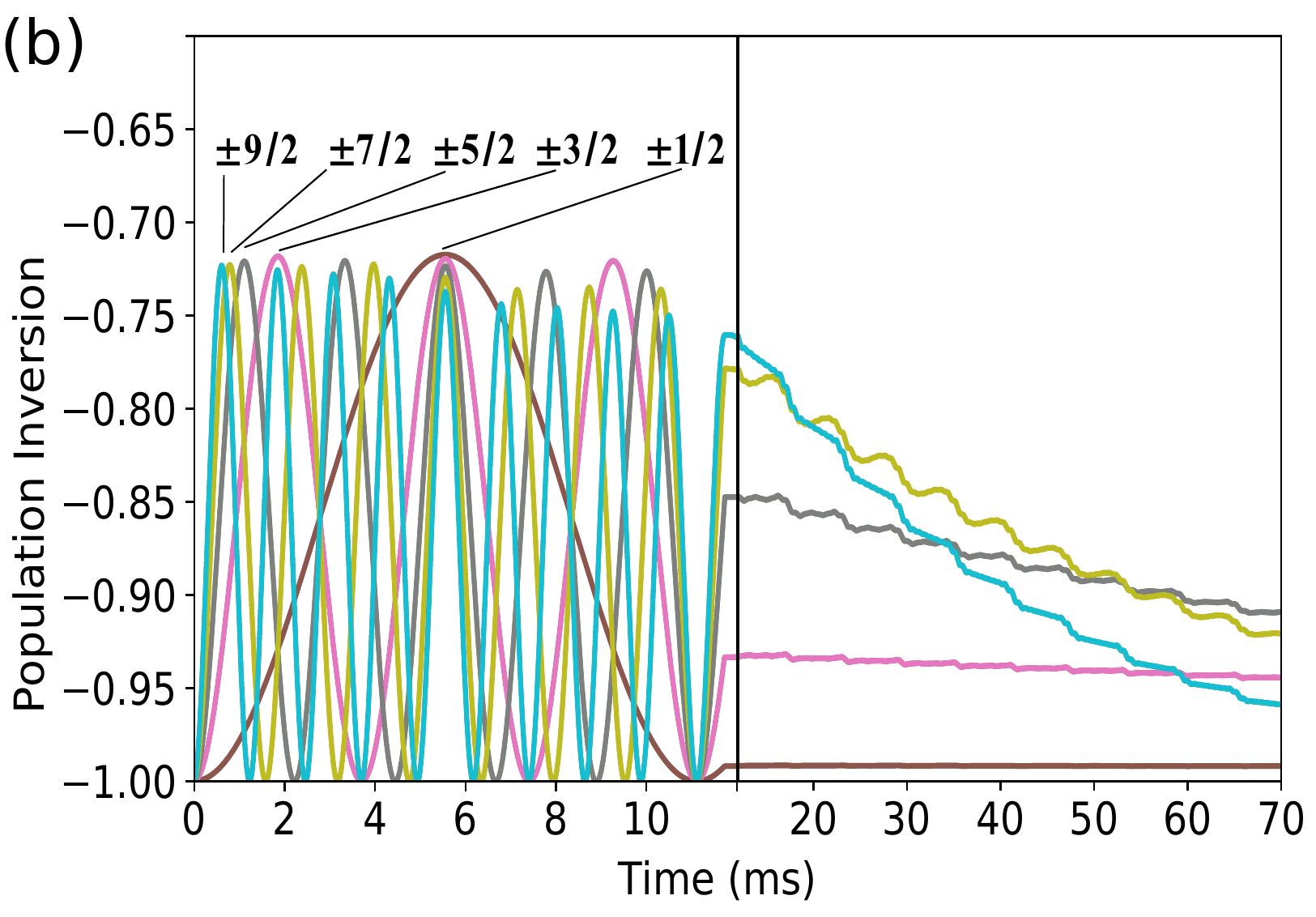}
\par\end{centering}
\caption{\label{fig:beats-ten-ensembles}Preparation, superradiant emission and heterodyne detection of $4\times10^{5}$ ${}^{87}{\rm Sr}$  atoms evenly distributed over ten two-level transitions with $m_{F}=\pm9/2,...,\pm1/2$. Panels (a) and (b) show the intra-cavity mean photon number and population difference among the excited and ground states for each value of $m_F$. The inset in panel (a) shows the simulated power spectral density (black noisy curve) obtained by a Fourier transform of the  simulated heterodyne detection signal, and Lorentzian fits to the spectral peaks (red solid curves). The results in panel (a) is in good agreement with the experimental results in Fig. 2(c) in \citep{MANorcia}.  The physical parameters of the calculations  are specified in the text.  }
\end{figure}

We apply the above stochastic mean-field equations to simulate the frequency measurements on the pulsed superradiant signals, as reported in \citep{MANorcia}. To obtain the results in Fig. \ref{fig:beats-ten-ensembles}, we assume the resonant condition  $\omega_c=\omega_a$ for the system without the magnetic field. A magnetic field of $1.53$ Gauss results in ten $\pi$-transitions with frequencies  $\omega_{\pm9/2},{}_{\pm7/2},{}_{\pm5/2},{}_{\pm3/2},{}_{\pm1/2}/2\pi=\omega_{a}/2\pi\pm750.00,583.33,416.67,250.00,83.33$ Hz (equal spacing of $166.7$ Hz). We drive the cavity resonantly (i.e. $\omega_{d}=\omega_{c}$) by a square laser pulse of duration $T=11.5$ ms and coupling strength $\Omega(t)=\Omega_m=2\pi\times7.5\times10^{3}$ $\sqrt{\mathrm{Hz}}$, and we then switch on the heterodyne detection, which relies on a reference laser blue shifted by  $\Delta=2\pi$ kHz  from the cavity mode, i.e. $\omega_l = \omega_{c} + \Delta$. We assume a detection efficiency of $\eta=0.12$ for our simulations \citep{MANorcia}. 

Fig. \ref{fig:beats-ten-ensembles} (a) shows that the intra-cavity photon number increases dramatically and oscillates around $4850$ when the driving laser is on,  and drops dramatically when the driving field is switched off, and finally yields a complex beats-pattern, in good agreement with Fig. 2(c) in the experimental article \citep{MANorcia}. Fig. \ref{fig:beats-ten-ensembles} (b) shows that the population difference $\bigl\langle\hat{\sigma}_{m_{F}}^{z}\bigr\rangle$ of different sub-ensembles follows oscillations of different frequencies in the preparation stage ($0$ ms to $11.5$ ms), and then decays monotonically and shows small ripples in the superradiant stage  ($11.5$ ms to $70$ ms).

These results can be understood by noting that for a classical laser pulse with amplitude $\sqrt{\kappa_{1}}\Omega_{m} \gg\kappa$, the field inside the cavity behaves like a classical field with an amplitude $\left\langle \hat{a}\right\rangle \approx i 2\sqrt{\kappa_{1}}\Omega_{m}/\kappa$ (for $\omega_c =\omega_d$). The optical Bloch equations for two-level atoms with transition frequency $\omega_0$ driven with the strength $\nu$ by a classical field of frequency $\omega_f$, yield a population in the excited state $P_{e}= \left|2\nu/\Omega\right|^2 {\rm sin}^2(\Omega T/2)$ at time $T$ with the Rabi oscillation frequency $\Omega =\sqrt{\left(\omega_f- \omega_0\right)^2 +(2\nu)^2 }$. By assuming the Zeeman shifted transition frequencies, $\omega_0 = \omega_{a} + \Delta_B m_F$, the optical frequency $\omega_f = \omega_c = \omega_a$, and the  $m_F$ dependent transition strength, $\nu=\left\langle \hat{a}\right\rangle m_{F} g_0/\sqrt{F\left(F+1\right)}$, we obtain   $\Omega = |m_F| \Omega_0$ with $\Omega_0 = \sqrt{\Delta_B^2  +|2\left\langle \hat{a}\right\rangle g_0/\sqrt{F\left(F+1\right)}|^2}$. It follows that   $\bigl\langle\hat{\sigma}_{m_{F}}^{z}\bigr\rangle = 2P_e - 1 = 2 P_e^{max} \mathrm{sin}^{2}\left(\left|m_{F}\right|\Omega_{0}T/2\right)-1$, and that the maximal excited-state population $P_e^{max} = \left|2\nu/\Omega\right|^2  =\left|2\left\langle \hat{a}\right\rangle g_{0}/\Omega_{0}\sqrt{F\left(F+1\right)}\right|^{2}$, does not depend on $m_F$ as it appears both in the numerator and the denominator. As a result, $\bigl\langle\hat{\sigma}_{m_{F}}^{z}\bigr\rangle$ oscillate with the same maximum value $2 P_e^{max}-1$ but different Rabi-frequencies $\left|m_{F}\right|\Omega_{0}$ for different sub-ensembles, see Fig. \ref{fig:beats-ten-ensembles}(b). Since these frequencies are proportional to $\left|m_{F}\right|$ and are also equally spaced,  $\bigl\langle\hat{\sigma}_{m_{F}}^{z}\bigr\rangle$  oscillates faster for the sub-ensembles with larger $m_{F}$ and arrives at the same maximum at integer multiples of the period $T=2\pi/\Omega_{0}$. To reproduce the superradiant pattern in the experiment \citep{MANorcia}, we choose $T$ such that the population difference decreases with reducing  $\left|m_{F}\right|$, see the curve crossings with the vertical line. In the superradiant stage, the population difference $\bigl\langle\hat{\sigma}_{m_{F}}^{z}\bigr\rangle$ decay faster for the sub-ensembles with larger $\left|m_{F}\right|$, reflecting
the larger Purcell enhanced atomic decay rate $\Gamma_{m_{F}}=4\left(m_{F}g_{0}/\sqrt{F\left(F+1\right)}\right)^{2}\kappa/\left[m_{F}^{2}\Delta_{B}^{2}+\kappa^{2}\right]$
for larger $\left|m_{F}\right|$ (notice $\left|m_{F}\right|\Delta_{B}\ll\kappa$). In addition, it also shows small steps caused by the constructive and destructive interference between the atomic subensembles emitting on different transitions (not shown).

The balanced heterodyne detection yields a fluctuating photon-current difference (see the Appendix  \ref{subsec:currentamplitude}), and the Fourier transform of this signal yields a power spectral density with eight peaks, which can be fitted by Lorentzian functions with the frequencies $(\omega_{9/2}-\omega_a)/2\pi=749.53$ Hz, $(\omega_{-9/2}-\omega_a)/2\pi=-750.60$ Hz, $(\omega_{7/2}-\omega_a)/2\pi=582.91$ Hz, $(\omega_{-7/2}-\omega_a)/2\pi=-582.53$
Hz, $(\omega_{5/2}-\omega_a)/2\pi=415.69$ Hz, $(\omega_{-5/2}-\omega_a)/2\pi=-416.41$ Hz, $(\omega_{3/2}-\omega_a)/2\pi=250.61$ Hz, $(\omega_{-3/2}-\omega_a)/2\pi=-251.76$ Hz, see the inset of Fig. \ref{fig:beats-ten-ensembles} (a). The transitions $m_F=\pm1/2$ with frequencies $\omega_{\pm1/2}$ couple too weakly with the cavity mode to be resolved.  The difference between the extracted and expected frequencies is caused by the detection noise and contributes to the frequency uncertainty of the scheme (see below).

\paragraph{Uncertainty of the Frequency Measurement}

\begin{figure}
\begin{centering}
\includegraphics[scale=0.45]{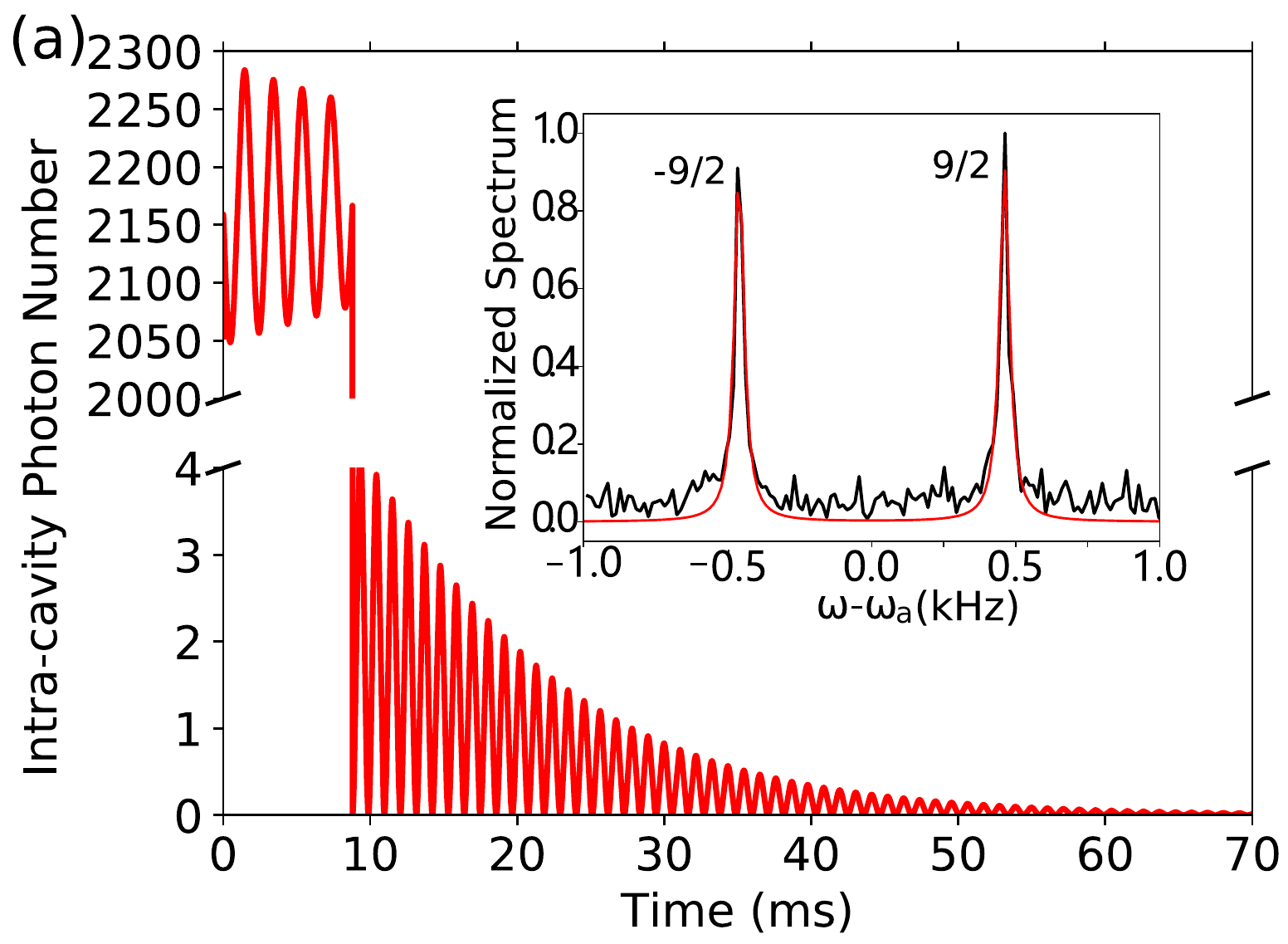}
\includegraphics[scale=0.45]{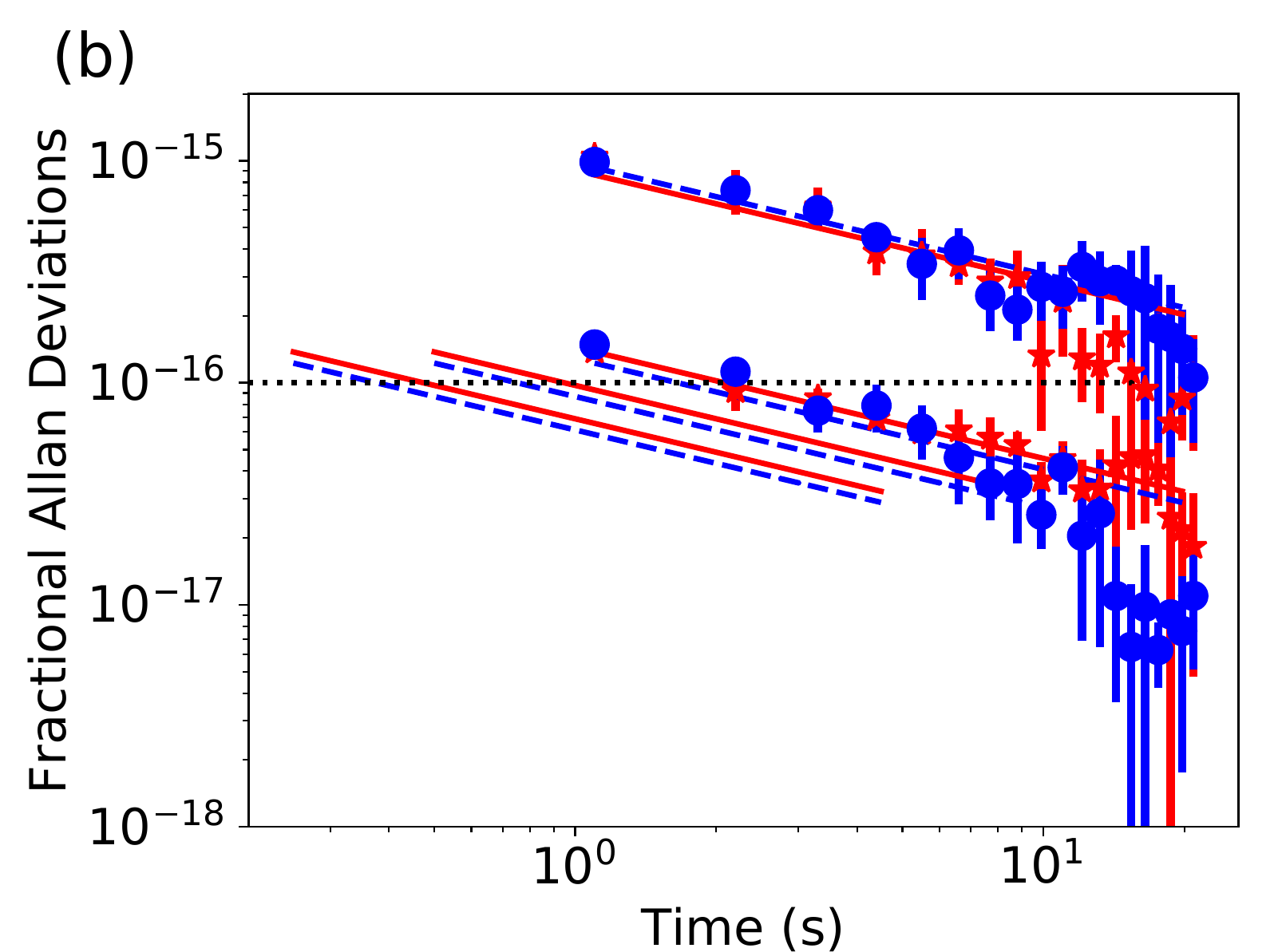}
\par\end{centering}
\caption{\label{fig:beats-two-ensembles} Frequency measurement on pulsed superradiant emission from
$1.8\times10^{5}$ ${}^{87}{\rm Sr}$ ${}^{87}{\rm Sr}$ atoms distributed evenly on the two extreme $m_{F}=\pm9/2$ atomic transitions.  Panel (a) shows the intra-cavity photon number, while the inset shows the power spectral density (relative to $\omega_a$, black noisy curve) and the fitting with two Lorentzian functions (red solid curves). Panel (b) shows the fractional Allan deviation $\sigma(\tau)$ for the center (blue dots) and difference (red squares) of the peak frequencies $\omega_{\pm9/2}$ after multiple repetitions of the simulated experiment, which each lasts for $T_{c}=1.1$ s \cite{MANorcia}. The mean and standard deviation are obtained by averaging the frequencies over three times forty independent simulations. The upper dots and squares are for the system considered in panel (a), while the lower ones are for optimized system parameters giving longer superradiance pulses. The results of the fitting of these data (solid and dashed lines) are explained in the text. The lower red solid (blue dashed) lines on the left side show the results for $T_{c}=0.5$ and $0.25$ s.  The horizontal dotted lines shows the noise floor $10^{-16}$ of the reference laser used in \cite{MANorcia}. Panel (a) agrees with Fig. 2(b) in the experimental article  \cite{MANorcia}, and the upper curves and dots in panel (b) are in good agreement with the upper line of Fig. 3(a) in \cite{MANorcia}. Other parameters are the same as used in Fig. \ref{fig:beats-ten-ensembles} or specified in the text. }   
\end{figure}

So far, we consider atoms distributed over all the $\pi$-transitions. It may be better to distribute the atoms only to the two extreme transitions  $m_F = \pm 9/2$ since they couple more strongly with the cavity mode and provide two symmetrically separated peaks in the power spectral density, which are sufficient to determine the intrinsic atomic transition frequency and the applied static magnetic field. Fig. \ref{fig:beats-two-ensembles} demonstrates the frequency measurement for such a system with $1.8\times10^{5}$  ${}^{87}{\rm Sr}$ atoms. There, a magnetic field of $0.94$ G results in the atomic transition frequencies $\omega_{\pm9/2}/2\pi=\omega_{a}/2\pi\pm0.46$
kHz, and the excitation laser pulse has a strength of $\Omega_{m}=2\pi\times5\times10^{3}$
$\sqrt{\mathrm{Hz}}$ and a duration of $8.8$ ms.  

Fig. \ref{fig:beats-two-ensembles} (a) shows that the intra-cavity photon number oscillates around the average value $2150$ and drops dramatically when the driving laser is switched off, leaving about $15\%$ of the atoms in the excited state (not shown). The system then shows superradiant beats and the power spectral density displays two peaks ( see the inset of Fig. \ref{fig:beats-two-ensembles} (a)),  which are consistent with Fig. 2(b) in the experiment  \citep{MANorcia}. The two peaks can be fitted with Lorentzian functions with the frequencies $(\omega_{-9/2}-\omega_a)/2\pi=-460.31$ Hz, $(\omega_{9/2}-\omega_a)/2\pi=459.91$ Hz and line-widths $12.31$ Hz, $14.03$ Hz. These frequencies deviate from the expected values by $0.31$ Hz and $0.09$ Hz. 

The uncertainty of the frequency measurement as a function of the total duration $\tau$ of the experiment is characterized by the so-called fractional Allan deviation $\sigma\left(\tau\right)$. To compute this quantity, let us assume that we need the time $T_c$ to carry out a single experiment, e.g.  $T_c= 1.1$ s in \cite{MANorcia}, and that we have carried out $N$ experiments to yield $N$ candidate frequencies $\bar{\omega}_{n}$. Then, $\sigma\left(\tau\right)$ for the probing time $\tau = T_c$ can be evaluated with the formula: $\sigma\left(\tau\right)=\sqrt{\frac{1}{N-1}\sum_{n=1}^{N-1} (\bar{\omega}_{n+1}-\bar{\omega}_{n})^{2} /(2\omega_{a}^{2})}$. Consistently with the strategy applied in the experiment  \citep{MANorcia}, we consider the average of $m$ frequencies  ($m < N/2$) as the frequency measured with a single experiment of total duration  $\tau=m T_c$, and we thus obtain $N/m$ average frequencies by which we compute $\sigma\left(\tau\right)$ for $\tau = m T_c$. 

Fig.\ref{fig:beats-two-ensembles} (b) shows the computed fractional Allan deviation $\sigma\left(\tau\right)$ for the center frequency $\omega_n = \left(\omega_{9/2}+\omega_{-9/2}\right)/2$ (blue dots) and the frequency difference $\omega_n = \left(\omega_{9/2}-\omega_{-9/2}\right)/2$ (red squares) from the two extracted frequencies $\omega_{\pm9/2}$. For the average frequency (frequency difference),  $\sigma\left(\tau\right)$ can be fitted with $9.06\times 10^{-16}/\sqrt{\tau/s}$ ($9.75\times 10^{-16}/\sqrt{\tau/s}$), in excellent agreement with the measured Allan deviation of the average frequency $\sigma\left(\tau\right) = 1.04\times 10^{-15}/\sqrt{\tau/s}$, but a factor $2.5$ larger than that measured for the frequency difference \cite{MANorcia}. Based on this observation, we may assume that our simulations yield the same or an upper bound of the Allan deviation of the frequency measurements. 

To optimize the frequency measurement, we notice that the duration of the superradiant pulse is inversely proportional to the number of atoms \cite{MANorcia-2} and it can be affected by the atomic initial states and the magnetic field, see the Appendix \ref{subsec:optimization}. Thus, by reducing the number of atoms to $ 9\times 10^4$ and assuming a smaller Zeeman splitting for the transition frequencies, $\omega_{\pm9/2} = \omega_a \pm 2\pi \times 50  {\rm Hz}$, and using a rectangle laser pulse of strength $10\times 10^3 \sqrt{{\rm Hz}}$ and shorter duration $T=1.1$ ms, we can prepare the atoms with more than $97\%$ population in the excited state, and obtain a $300$ ms long superradiant-beat signal with two frequency peaks about $50$ Hz from the intrinsic atomic transition frequency, see the Appendix \ref{subsec:optimizedradiation}. These peaks lead to a fractional Allan deviation  $1.44,1.29\times 10^{-16}/\sqrt{\tau/s}$ for the center frequency and the frequency difference, respectively,  see Fig. \ref{fig:beats-two-ensembles} (b). 

Before we can achieve the predicted fractional Allan deviation  $\sigma\left(\tau\right)$, we must address the noise of the reference laser and the loss of atoms during the measurements.  In the experiment \citep{MANorcia}, the reference laser is stabilized with a high-quality optical cavity, which has a thermal noise floor of $1\times10^{-16}$, see the horizontal dashed line in Fig. \ref{fig:beats-two-ensembles} (b), and new atoms must be reloaded into the cavity before the optical excitation and measurement is carried out, which in total takes about $T_{c}=1.1$ s or longer. To overcome these obstacles, we can reduce the noise level of the reference laser by stabilizing it to an ultra high-quality optical cavity, and we may compensate the atom loss by continuously injecting new atoms \citep{CCChen}, see the Appendix \ref{subsec:atomlosspumping}. If we can reduce the time for single experiments to, e.g., $T_{c}=0.5$ s or $0.25$ s, $\sigma\left(\tau\right)$ can be reduced to $9\times10^{-17}/ \sqrt{\tau/s}$ or $7\times 10^{-17}/\sqrt{\tau/s}$, as indicated with the shifted blue dashed and red solid lines in Fig. \ref{fig:beats-two-ensembles}(b), which is comparable with the record of $6\times 10^{-17}/\sqrt{\tau/s}$ \citep{MSchioppo}.  

In the above simulations, the atomic sub-ensembles are initially prepared in superposition states, where the atoms have an initial polarization, $\left \langle \hat{\sigma}_i^- \right \rangle \neq 0$. According to the first line of Eq. (\ref{eq:amplitude}), this polarization feeds the cavity field amplitude, $\left \langle \hat{a} \right \rangle \neq 0$, and drives the photon-current in the heterodyne detection, see Eq. (\ref{current}), allowing us to detect the frequencies. For atoms prepared in the fully excited state, there is no initial mean polarization and a classical mean field argument suggests that it would not be possible to observe the frequencies with heterodyne detection. This argument, however, is incorrect. We show in the Appendix \ref{subsec:comparison} that the backaction of heterodyne detection causes a breaking of the symmetry and establishes optical coherence in the system, see the second line of Eq. (\ref{eq:amplitude}), and thus yields a modulated heterodyne current even in the case with no initial coherence.

\paragraph{Conclusion}

In summary, we have combined the theory of continuous quantum measurements and a mean field description of cavity-QED with many emitters to simulate heterodyne detection of pulsed superradiance from tens of thousands of ${}^{87}{\rm Sr}$ atoms trapped in a one-dimensional optical lattice inside an optical cavity. Our simulations show that the computed frequency uncertainty decreases as $\sim 9\times10^{-16}/\sqrt{\tau/s}$ with the measurement time $\tau$, in agreement with recent experiments \cite{MANorcia}.  By use of longer superradiance pulses and a shorter duty cycle for each single measurement, the frequency uncertainty may be further reduced by one to two orders of magnitude and thus becomes comparable with current records \cite{MSchioppo}. 

Our theory can be directly applied to the study of frequency measurements on superradiant pulses from other alkaline-earth atoms, such as strontium-88 atoms \citep{MANorcia1,SASchaffer} and  calcium atoms \citep{TLaske}. Steady-state superradiance \citep{DMeiser} and superradiant Raman lasers \citep{JMWeiner} may be subject to similar analyses, which may also reveal more exotic effects of quantum measurements such as conditional entanglement and spin squeezing in the optical lattice clock systems.

\begin{acknowledgments}
The authors thank James K. Thompson for generously sharing details and insights from the experiments on superradiance frequency measurements \citep{MANorcia,MANorcia-2}. This work was supported by the National Natural Science Foundation of China through the project No. 12004344 and the Danish National Research Foundation through the Grant Agreement No. DNRF156. 
\end{acknowledgments}

\newpage
\appendix

\section{Second-order Mean-field Equations \label{sec:observables}}

In the main text, we have presented the stochastic master equation
(\ref{eq:stochastic-master-equation}) for the heterodyne detection of 
superradiance from strontium atoms and have outlined the procedure to solve it with the second-order mean-field
theory. There, we have shown the equations for the intra-cavity photon number $\left\langle \hat{a}^{+}\hat{a}\right\rangle $ and the cavity field amplitude $\left\langle \hat{a}\right\rangle $,
see Eqs. (\ref{eq:photon-number}) and (\ref{eq:amplitude}).

In the following, we present the equations for other quantities. The equations for the atomic polarization
$\left\langle \hat{\sigma}_{i}^{-}\right\rangle $ and the population difference
$\left\langle \hat{\sigma}_{i}^{z}\right\rangle $ read
\begin{align}
 & \frac{\partial}{\partial t}\left\langle \hat{\sigma}_{i}^{-}\right\rangle =-i\tilde{\omega}_{i}\left\langle \hat{\sigma}_{i}^{-}\right\rangle +ig_{i}\left\langle \hat{a}\hat{\sigma}_{i}^{z}\right\rangle \nonumber \\
 & +\frac{dW}{dt}\sqrt{\eta\kappa_2}\Bigl[e^{i\Delta t}\left(\left\langle \hat{a} \hat{\sigma}_{i}^{+} \right\rangle^* -\left\langle \hat{a}\right\rangle ^{*}\left\langle \hat{\sigma}_{i}^{-} \right\rangle \right)\nonumber \\
 & +e^{-i\Delta t}\left(\left\langle  \hat{a} \hat{\sigma}_{i}^{-}  \right\rangle -\left\langle \hat{\sigma}_{i}^{-} \right\rangle \left\langle \hat{a}\right\rangle \right)\Bigr],  
\end{align}
\begin{align}
 & \frac{\partial}{\partial t}\left\langle \hat{\sigma}_{i}^{z} \right\rangle =4g_{i}\mathrm{Im}\left\langle \hat{a}\hat{\sigma}_{i}^{+}\right\rangle  -\gamma_{i}\left(1+\left\langle \hat{\sigma}_{i}^{z}\right\rangle \right) \nonumber \\
 & +\frac{dW}{dt}\sqrt{\eta\kappa_2}\Bigl[e^{i\Delta t}\left(\left\langle \hat{a} \hat{\sigma}_{i}^{z} \right\rangle^* -\left\langle \hat{a}\right\rangle ^{*}\left\langle \hat{\sigma}_{i}^{z} \right\rangle \right)\nonumber \\
 & +e^{-i\Delta t}\left(\left\langle \hat{a} \hat{\sigma}_{i}^{z}  \right\rangle -\left\langle \hat{\sigma}_{i}^{z} \right\rangle \left\langle \hat{a}\right\rangle \right)\Bigr].
\end{align}
 Here, we have introduced the complex frequency $\tilde{\omega}_{i}=\omega_{i}-i\gamma_{i}/2$.
The equations for the atom-atom correlations read
\begin{align}
 & \frac{\partial}{\partial t}\left\langle \hat{\sigma}_{j}^{-}\hat{\sigma}_{i}^{+}\right\rangle =-i\left(\tilde{\omega}_{j}-\tilde{\omega}_{i}^{*}\right)\left\langle \hat{\sigma}_{j}^{-}\hat{\sigma}_{i}^{+}\right\rangle \nonumber \\
 & +ig_{j}\left\langle \hat{a}\hat{\sigma}_{j}^{z}\hat{\sigma}_{i}^{+}\right\rangle -ig_{i}\left\langle \hat{a}^{+}\hat{\sigma}_{j}^{-}\hat{\sigma}_{i}^{z}\right\rangle \nonumber \\
 & +\frac{dW}{dt}\sqrt{\eta\kappa_2}\Bigl[e^{i\Delta t}\left(\left\langle \hat{a}^{+} \hat{\sigma}_{j}^{-}\hat{\sigma}_{i}^{+} \right\rangle -\left\langle \hat{a}\right\rangle ^{*}\left\langle \hat{\sigma}_{j}^{-}\hat{\sigma}_{i}^{+} \right\rangle \right)\nonumber \\
 & +e^{-i\Delta t}\left(\left\langle \hat{\sigma}_{j}^{-}\hat{\sigma}_{i}^{+} \hat{a}  \right\rangle -\left\langle \hat{\sigma}_{j}^{-}\hat{\sigma}_{i}^{+} \right\rangle \left\langle \hat{a}\right\rangle \right)\Bigr], 
\end{align}
\begin{align}
 & \frac{\partial}{\partial t}\left\langle \hat{\sigma}_{j}^{-}\hat{\sigma}_{i}^{-}\right\rangle =-i\left(\tilde{\omega}_{i}+\tilde{\omega}_{j}\right)\left\langle \hat{\sigma}_{j}^{-}\hat{\sigma}_{i}^{-}\right\rangle \nonumber \\
 & +ig_{i}\left\langle \hat{a}\hat{\sigma}_{j}^{-}\hat{\sigma}_{i}^{z}\right\rangle +ig_{j}\left\langle \hat{a}\hat{\sigma}_{j}^{z}\hat{\sigma}_{i}^{-}\right\rangle \nonumber \\
 & +\frac{dW}{dt}\sqrt{\eta\kappa_2}\Bigl[e^{i\Delta t}\left(\left\langle \hat{a}^{+} \hat{\sigma}_{j}^{-}\hat{\sigma}_{i}^{-} \right\rangle -\left\langle \hat{a}\right\rangle ^{*}\left\langle \hat{\sigma}_{j}^{-}\hat{\sigma}_{i}^{-} \right\rangle \right)\nonumber \\
 & +e^{-i\Delta t}\left(\left\langle \hat{\sigma}_{j}^{-}\hat{\sigma}_{i}^{-} \hat{a}  \right\rangle -\left\langle \hat{\sigma}_{j}^{-}\hat{\sigma}_{i}^{-} \right\rangle \left\langle \hat{a}\right\rangle \right)\Bigr], 
\end{align}
\begin{align}
 & \frac{\partial}{\partial t}\left\langle \hat{\sigma}_{j}^{-}\hat{\sigma}_{i}^{z}\right\rangle =-i\tilde{\omega}_{j}\left\langle \hat{\sigma}_{j}^{-}\hat{\sigma}_{i}^{z}\right\rangle +ig_{j}\left\langle a\hat{\sigma}_{j}^{z}\hat{\sigma}_{i}^{z}\right\rangle \nonumber \\
 & -i2g_{i}\left[\left\langle \hat{a}\hat{\sigma}_{j}^{-}\hat{\sigma}_{i}^{+}\right\rangle -\left\langle \hat{a}^{+}\hat{\sigma}_{j}^{-}\hat{\sigma}_{i}^{-}\right\rangle \right]\nonumber \\
 & -\gamma_{i}\left(\left\langle \hat{\sigma}_{j}^{-}\right\rangle +\left\langle \hat{\sigma}_{j}^{-}\hat{\sigma}_{i}^{z}\right\rangle \right) \nonumber \\
 & +\frac{dW}{dt}\sqrt{\eta\kappa_2}\Bigl[e^{i\Delta t}\left(\left\langle \hat{a}^{+} \hat{\sigma}_{j}^{-}\hat{\sigma}_{i}^{z} \right\rangle -\left\langle \hat{a}\right\rangle ^{*}\left\langle \hat{\sigma}_{j}^{-}\hat{\sigma}_{i}^{z} \right\rangle \right)\nonumber \\
 & +e^{-i\Delta t}\left(\left\langle \hat{\sigma}_{j}^{-}\hat{\sigma}_{i}^{z} \hat{a}  \right\rangle -\left\langle \hat{\sigma}_{j}^{-}\hat{\sigma}_{i}^{z} \right\rangle \left\langle \hat{a}\right\rangle \right)\Bigr], 
\end{align}
\begin{align}
 & \frac{\partial}{\partial t}\left\langle \hat{\sigma}_{j}^{z}\hat{\sigma}_{i}^{z}\right\rangle =4g_{i}\mathrm{Im}\left\langle \hat{a}\hat{\sigma}_{j}^{z}\hat{\sigma}_{i}^{+}\right\rangle +4g_{j}\mathrm{Im}\left\langle a\hat{\sigma}_{j}^{+}\hat{\sigma}_{i}^{z}\right\rangle \nonumber \\
 & -\gamma_{i}\left(\left\langle \hat{\sigma}_{j}^{z}\right\rangle +\left\langle \hat{\sigma}_{j}^{z}\hat{\sigma}_{i}^{z}\right\rangle \right)-\gamma_{j}\left(\left\langle \hat{\sigma}_{i}^{z}\right\rangle +\left\langle \hat{\sigma}_{j}^{z}\hat{\sigma}_{i}^{z}\right\rangle \right) \nonumber \\
 & +\frac{dW}{dt}\sqrt{\eta\kappa_2}\Bigl[e^{i\Delta t}\left(\left\langle \hat{a}^{+} \hat{\sigma}_{j}^{z}\hat{\sigma}_{i}^{z} \right\rangle -\left\langle \hat{a}\right\rangle ^{*}\left\langle \hat{\sigma}_{j}^{z}\hat{\sigma}_{i}^{z} \right\rangle \right)\nonumber \\
 & +e^{-i\Delta t}\left(\left\langle \hat{\sigma}_{j}^{z}\hat{\sigma}_{i}^{z} \hat{a}  \right\rangle -\left\langle \hat{\sigma}_{j}^{z}\hat{\sigma}_{i}^{z} \right\rangle \left\langle \hat{a}\right\rangle \right)\Bigr].  \label{eq:two-atom-inversion}
\end{align}
 The equation for the photon-photon correlation reads
\begin{align}
 & \frac{\partial}{\partial t}\left\langle \hat{a}^{2}\right\rangle =-2i\tilde{\omega}_{c}\left\langle \hat{a}^{2}\right\rangle -i2\sqrt{\kappa_{1}}\Omega^{*}\left(t\right)e^{-i\omega_{d}t}\left\langle \hat{a}\right\rangle \nonumber \\
 & -2i\sum_{i}g_{i}N_{i}\left\langle \hat{a} \hat{\sigma}_{i}^{-}\right\rangle +\frac{dW}{dt}\sqrt{\eta\kappa_2}\Bigl[e^{i\Delta t}\nonumber \\
 & \times \left(\left\langle \hat{a}^{+}\hat{a}^{2}\right\rangle -\left\langle \hat{a}\right\rangle ^{*}\left\langle \hat{a}^{2}\right\rangle \right)+e^{-i\Delta t}\left(\left\langle \hat{a}^{3}\right\rangle -\left\langle \hat{a}^{2}\right\rangle \left\langle \hat{a}\right\rangle \right)\Bigr].
\end{align}
The equations for the atom-photon correlations read
\begin{align}
 & \frac{\partial}{\partial t}\left\langle \hat{a}\hat{\sigma}_{i}^{+}\right\rangle =i\left(\tilde{\omega}_{i}^{*}-\tilde{\omega}_{c}\right)\left\langle \hat{a}\hat{\sigma}_{i}^{+}\right\rangle -i\sqrt{\kappa_{1}}\Omega^{*}\left(t\right)e^{-i\omega_{d}t}\left\langle \hat{\sigma}_{i}^{-}\right\rangle ^{*}\nonumber \\
 & -i\left(g_{i}/2\right)\left(1-\left\langle \hat{\sigma}_{i}^{z}\right\rangle \right)-ig_{i}\left(N_{i}-1\right)\left\langle \hat{\sigma}_{i}^{-}\hat{\sigma}_{i}^{+}\right\rangle \nonumber \\
 & -i\sum_{j\neq i}g_{j}N_{j}\left\langle \hat{\sigma}_{j}^{-}\hat{\sigma}_{i}^{+}\right\rangle -ig_{i}\left\langle \hat{a}\hat{a}^{+}\hat{\sigma}_{i}^{z}\right\rangle \nonumber \\
 & +\frac{dW}{dt}\sqrt{\eta\kappa_2}\Bigl[e^{i\Delta t}\left(\left\langle \hat{a}^{+}\hat{a}\hat{\sigma}_{i}^{+}\right\rangle -\left\langle \hat{a}\right\rangle ^{*}\left\langle \hat{a}\hat{\sigma}_{i}^{+}\right\rangle \right)\nonumber \\
 & +e^{-i\Delta t}\left(\left\langle \hat{a}\hat{a}\hat{\sigma}_{i}^{+}\right\rangle -\left\langle \hat{a}\hat{\sigma}_{i}^{+}\right\rangle \left\langle \hat{a}\right\rangle \right)\Bigr],
\end{align}
\begin{align}
 & \frac{\partial}{\partial t}\left\langle \hat{a}\hat{\sigma}_{i}^{-}\right\rangle =-i\left(\tilde{\omega}_{c}+\tilde{\omega}_{i}\right)\left\langle \hat{a}\hat{\sigma}_{i}^{-}\right\rangle -i\sqrt{\kappa_{1}}\Omega^{*}\left(t\right)e^{-i\omega_{d}t}\left\langle \hat{\sigma}_{i}^{-}\right\rangle \nonumber \\
 & -i\sum_{j\neq i}g_{j}N_{j}\left\langle \hat{\sigma}_{j}^{-}\hat{\sigma}_{i}^{-}\right\rangle -ig_{i}\left[\left(N_{i}-1\right)\left\langle \hat{\sigma}_{i}^{-}\hat{\sigma}_{i}^{-}\right\rangle -\left\langle \hat{a}^{2}\hat{\sigma}_{i}^{z}\right\rangle \right]\nonumber \\
 & +\frac{dW}{dt}\sqrt{\eta\kappa_2}\Bigl[e^{i\Delta t}\left(\left\langle \hat{a}^{+}\hat{a}\hat{\sigma}_{i}^{-}\right\rangle -\left\langle \hat{a}\right\rangle ^{*}\left\langle \hat{a}\hat{\sigma}_{i}^{-}\right\rangle \right)\nonumber \\
 & +e^{-i\Delta t}\left(\left\langle \hat{a}\hat{a}\hat{\sigma}_{i}^{-}\right\rangle -\left\langle \hat{a}\hat{\sigma}_{i}^{-}\right\rangle \left\langle \hat{a}\right\rangle \right)\Bigr],
\end{align}
\begin{align}
 & \frac{\partial}{\partial t}\left\langle \hat{a}\hat{\sigma}_{i}^{z}\right\rangle =-i\tilde{\omega}_{c}\left\langle \hat{a}\hat{\sigma}_{i}^{z}\right\rangle -i\sqrt{\kappa_{1}}\Omega^{*}\left(t\right)e^{-i\omega_{d}t}\left\langle \hat{\sigma}_{i}^{z}\right\rangle \nonumber \\
 & -ig_{i}\left[\left\langle \hat{\sigma}_{i}^{-}\right\rangle +\left(N_{i}-1\right)\left\langle \hat{\sigma}_{i}^{-}\hat{\sigma}_{i}^{z}\right\rangle \right]\nonumber \\
 & -i\sum_{j\neq i}g_{j}N_{j}\left\langle \hat{\sigma}_{j}^{-}\hat{\sigma}_{i}^{z}\right\rangle -i2g_{i}\left[\left\langle \hat{a}^{2}\hat{\sigma}_{i}^{+}\right\rangle -\left\langle \hat{a}\hat{a}^{+}\hat{\sigma}_{i}^{-}\right\rangle \right]\nonumber \\
 & -\gamma_{i}\left(\left\langle \hat{a}\right\rangle +\left\langle \hat{a}\hat{\sigma}_{i}^{z}\right\rangle \right)\nonumber \\
 & +\frac{dW}{dt}\sqrt{\eta\kappa_2}\Bigl[e^{i\Delta t}\left(\left\langle \hat{a}^{+}\hat{a}\hat{\sigma}_{i}^{z}\right\rangle -\left\langle \hat{a}\right\rangle ^{*}\left\langle \hat{a}\hat{\sigma}_{i}^{z}\right\rangle \right)\nonumber \\
 & +e^{-i\Delta t}\left(\left\langle \hat{a}\hat{a}\hat{\sigma}_{i}^{z}\right\rangle -\left\langle \hat{a}\hat{\sigma}_{i}^{z}\right\rangle \left\langle \hat{a}\right\rangle \right)\Bigr].
\end{align}
To close the equations, we apply the third order cumulant expansion
to approximate the third order correlations, e.g. $\left\langle \hat{a}\hat{\sigma}_{j}^{z}\hat{\sigma}_{i}^{+}\right\rangle$, with product of lower order correlations, e.g. $\left\langle \hat{a}\right\rangle \left\langle \hat{\sigma}_{j}^{z}\hat{\sigma}_{i}^{+}\right\rangle + 
\left\langle \hat{\sigma}_{j}^{z} \right\rangle \left\langle \hat{a} \hat{\sigma}_{i}^{+}\right\rangle+ \left\langle \hat{\sigma}_{i}^{+}\right\rangle \left\langle \hat{a} \hat{\sigma}_{j}^{z}\right\rangle -2 \left\langle \hat{a}\right\rangle \left\langle \hat{\sigma}_{j}^{z}\right\rangle \left\langle \hat{\sigma}_{i}^{+}\right\rangle$.

\section{Supplemental Numerical Results \label{sec:further-results}}
In this section, we provide more details on the  results of the numerical calculations than presented in the main text.

\begin{figure}[h]
\begin{centering}
\includegraphics[scale=0.43]{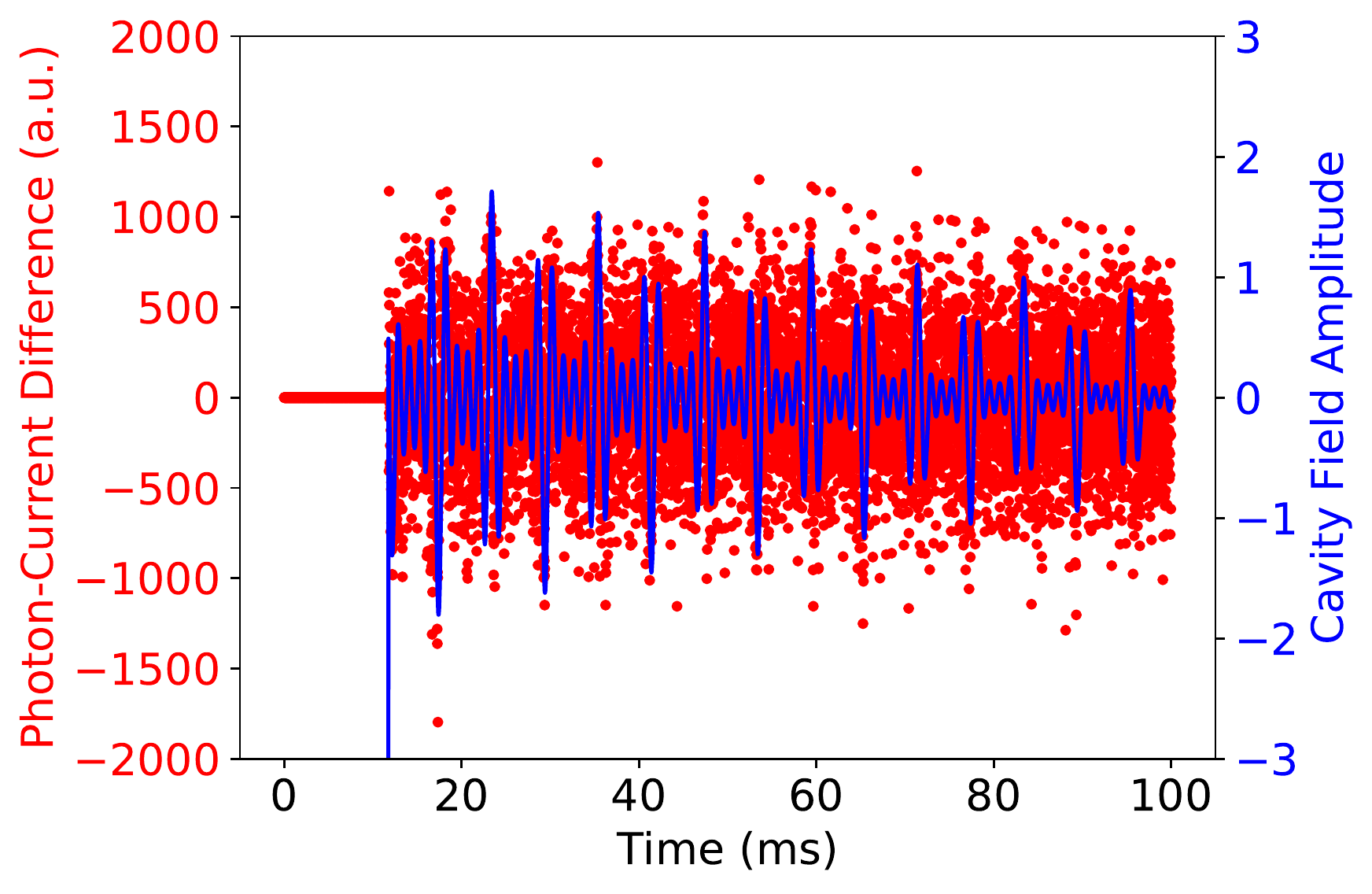}
\par\end{centering}
\caption{\label{fig:currentamplitude} Photon-current difference (red dots, left axis) and the cavity field amplitude (blue solid line, right axis) for the system studied in Fig. \ref{fig:beats-ten-ensembles}.} 
\end{figure}

\subsection{Photon-current Difference and Cavity Field Amplitude \label{subsec:currentamplitude}}
In Fig. \ref{fig:beats-ten-ensembles} in the main text, we presented the radiation pattern, the power spectral density and the population difference of  atomic sub-ensembles for a system with $4\times 10^5$ strontium atoms in subject to a static magnetic field. In Fig. \ref{fig:currentamplitude} we supplement these results with the photon-current difference  (red dots) and the cavity field amplitude (blue solid line). Here, we focus on the results in the superradiant stage (from $11.5$ ms to $100$ ms). We have verified that the cavity field amplitude is pure imaginary and the square of its module reproduces the cavity photon pattern as given in  Fig. \ref{fig:beats-ten-ensembles}. In contrast, the photon-current difference is rather noisy since it is affected by the photon short noise.  The Fourier transform of this noisy signal results to the power spectral density shown in the inset of Fig.  \ref{fig:beats-ten-ensembles}.  

\subsection{Superradiance for Different Initial Conditions and Magnetic Fields \label{subsec:optimization}}
\begin{figure}[htbp]
\begin{centering}
\includegraphics[scale=0.65]{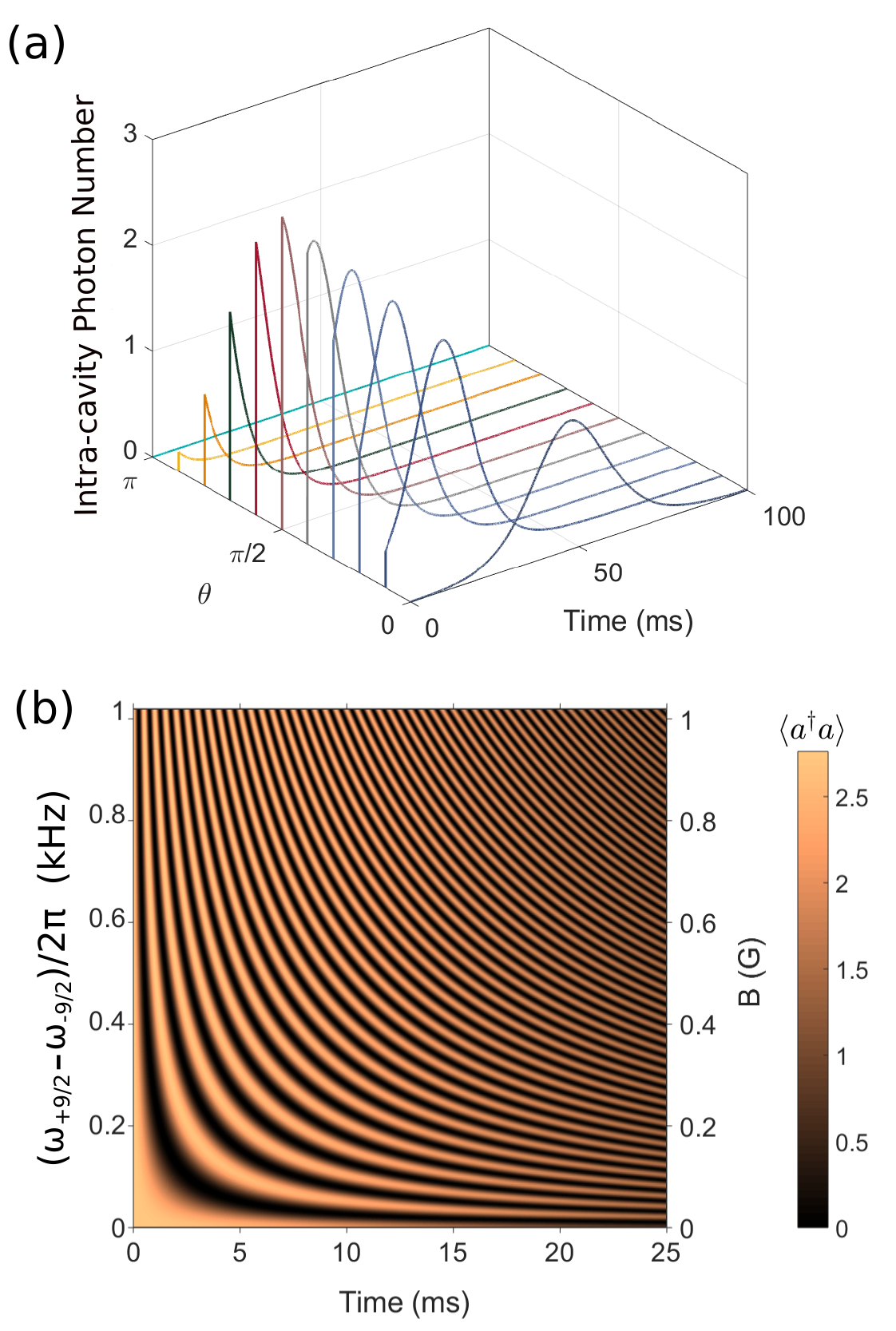}
\par\end{centering}
\caption{\label{fig:optimization}
Influence of the atomic initial states (a) and the magnetic field induced level splitting (b) on the superradiant signals from  $\sum_{i}N_{i}=1.8\times10^{5}$ atoms. In panel (a), the magnetic field vanishes and the initial states are characterized by the angle $\theta$ varying from $\pi$ (ground state) to $0$ (excited state). Panel (b) shows the time-dependent photon number in the cavity as function of time and frequency difference $\omega_{+9/2}-\omega_{-9/2}$. In (b), the atoms are initially prepared with even population in the ground and excited states. Other parameters see text. }
\end{figure}

In Fig. \ref{fig:beats-two-ensembles} in the main text, we have shown that the computed fractional Allan deviation for one system agrees qualitatively with the result reported in the experiment \citep{MANorcia}. To achieve much smaller fractional Allan deviation, i.e. smaller frequency uncertainty, we study in Fig. \ref{fig:optimization} how the atomic initial states (a) and the magnetic field (b) affect the superradiant signals. As revealed in the main text, the population of the atomic states follows the Rabi-oscillation in the preparation stage. Therefore, after the preparation, the  atomic states can be written as $\left|\psi\right\rangle =\mathrm{cos}\left(\theta/2\right)\left|u\right\rangle +\mathrm{\sin}\left(\theta/2\right)\left|l\right\rangle $ with the upper $\left|u\right\rangle$ and lower atomic state $\left|l\right\rangle$  (of the vertical transitions $m_F = \pm 9/2$). When the angle $\theta$ varies from $0$ to $\pi$, the atomic state changes from the excited state to the ground state. Fig. \ref{fig:optimization} (a) shows a $100$ ms long superradiant pulse centered around $50$ ms for the case where the atoms are initially fully excited. The pulse shifts to earlier time for states with less initial population in the excited states, and for $\theta<\pi/2$, i.e., the states with the majority population on the ground states, the signal decays monotonically with time. Fig. \ref{fig:optimization} (b) shows that for the atoms initially prepared in the state with even population in the ground and excited state ($\theta=\pi/2$) the superradiance forms beats within the first about $20$ ms if the magnetic field is stronger than $0.05$ Gauss, implying a frequency separation larger than $5$ Hz.

\subsection{Atomic Population Difference, Superradiance Pulse and Power Spectral Density for Optimized System \label{subsec:optimizedradiation}}

\begin{figure}[htbp]
\begin{centering}
\includegraphics[scale=0.45]{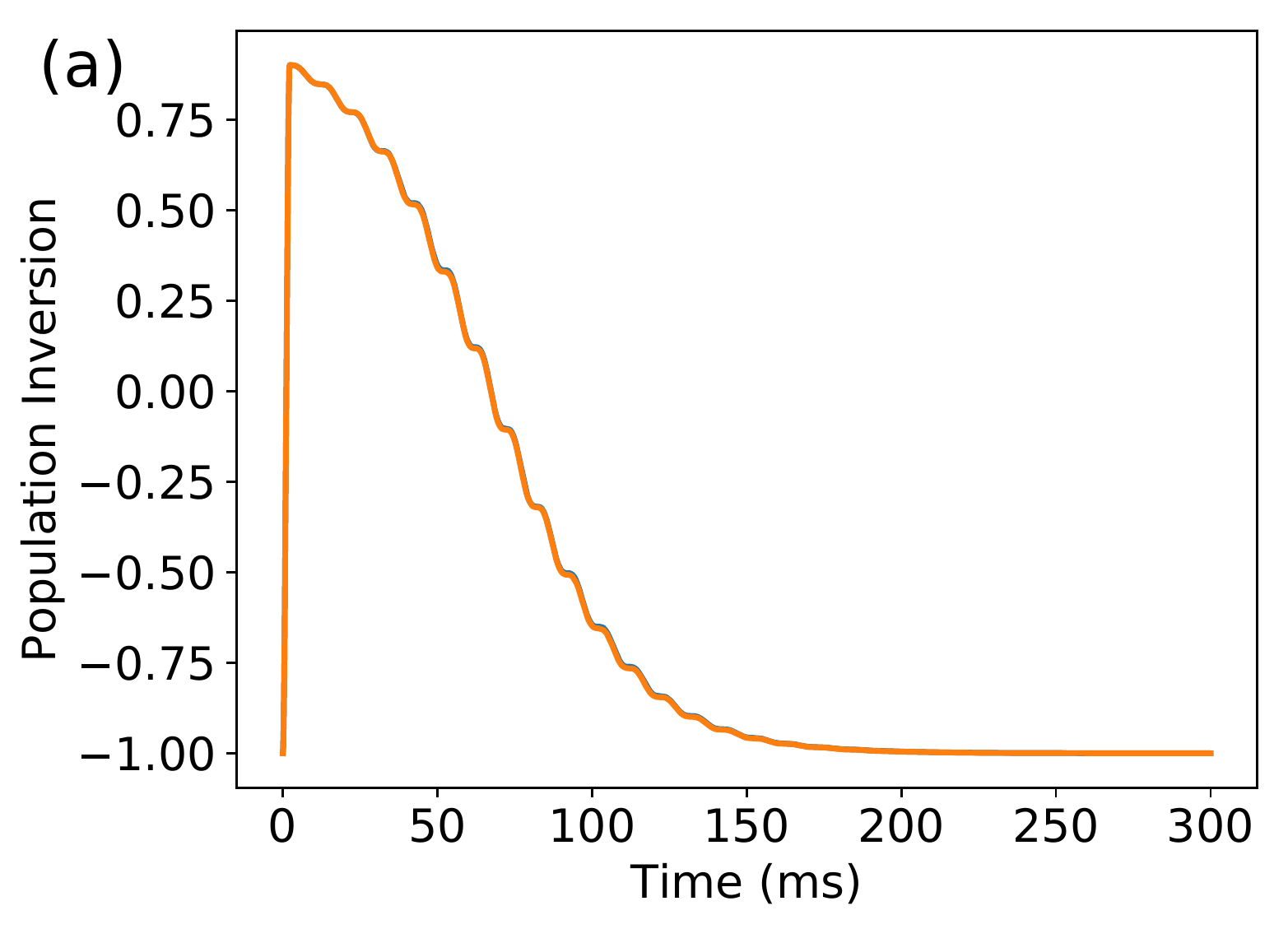}
\par\end{centering}
\begin{centering}
\includegraphics[scale=0.45]{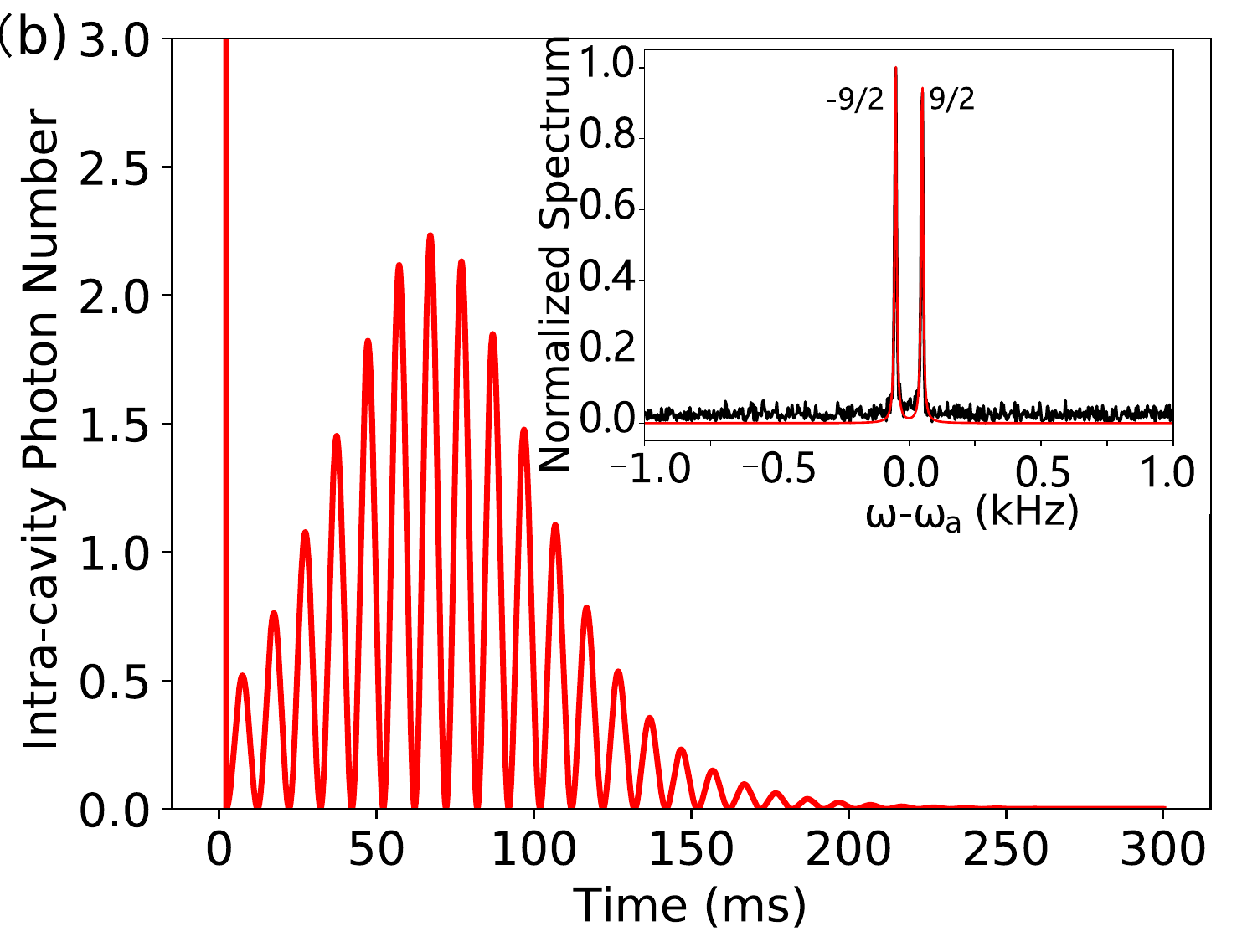}
\par\end{centering}
\caption{\label{fig:optimizedradiation}  Results for systems with $\sum_i N_i =  9 \times10^{4}$  ${}^{87}{\rm Sr}$ atoms, evenly distributed in two sub-ensembles with the transition frequencies $\omega_{\pm9/2} = \omega_a \pm 2\pi \times 50 {\rm Hz}$, and a cavity, driven by laser pulses of strength $\Omega_m/2\pi=10\times 10^3 \sqrt{{\rm Hz}}$ and duration $T=1.1$ ms. Panel (a) shows  the (identical) population difference  of the two atom-sub-ensembles. Panel (b) shows the intra-cavity photon number and the power spectral density (inset) of a single simulation.  Other parameters see text.} 
\end{figure}

In Fig. \ref{fig:beats-two-ensembles} in main text, we have analyzed the fractional Allan deviations for an optimized system supporting longer superradiance signal. In Fig. \ref{fig:optimizedradiation}, we complement these results with the population difference of the two atomic sub-ensembles (a), the superradiance pulses and the power spectral density (b).  Fig. \ref{fig:optimizedradiation}(a) shows that the population difference reaches the maximum of $0.95$ after the preparation stage (1.1 ms) and then decays with time, which is also decorated with small ripples. Fig. \ref{fig:optimizedradiation}(b) shows that the intra-cavity photon number oscillates in time and the envelop of the oscillations increases firstly, reaches the maximum around $75$ ms and finally decreases. The superradiance pulse lasts over $210$ ms, which is about three times longer than the pulse in Fig. \ref{fig:beats-two-ensembles}(a). The longer signal leads to two narrow peaks around $\pm 50$ Hz in the power spectral density [insert of Fig. \ref{fig:optimizedradiation}(b)]. 

\subsection{Superradiance Pulses from System with Atom Loss and Injection \label{subsec:atomlosspumping}}

\begin{figure}[htbp]
\begin{centering}
\includegraphics[scale=0.43]{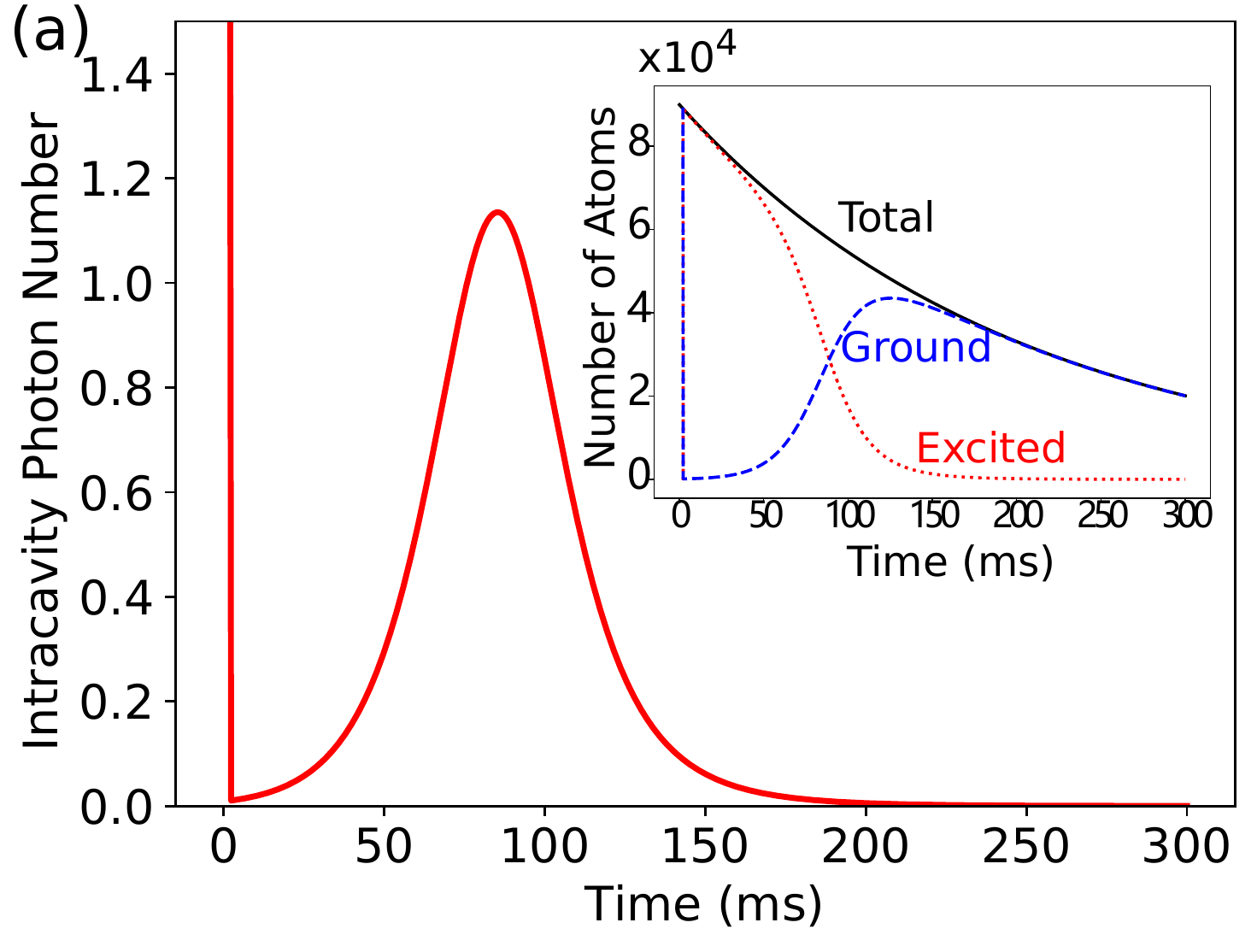}
\includegraphics[scale=0.43]{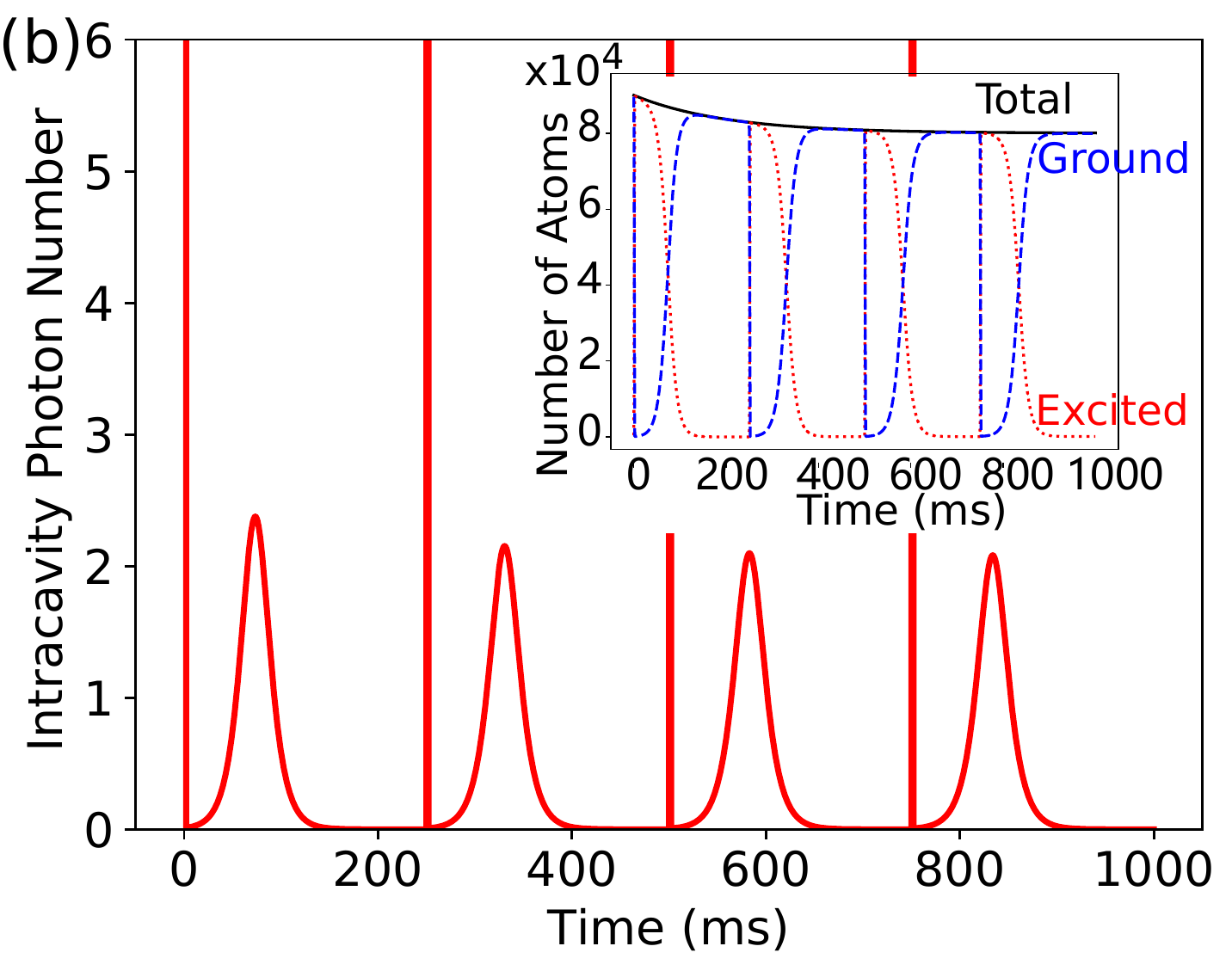}
\par\end{centering}
\caption{\label{fig:atomlosspumping}  Similar results as Fig. \ref{fig:optimizedradiation} except for the absence of the static magnetic field.  Panel (a) shows the intra-cavity photon number (red solid line), and the inset shows the total number of atoms $N(t)$ (black solid line), the number of excited [$\frac{1}{2}N(t)(1+\left\langle \hat{\sigma}_z \right \rangle(t))$,red dotted line] and ground [$\frac{1}{2}N(t)(1-\left\langle \hat{\sigma}_z \right\rangle (t))$, blue dashed line] atoms for  the system suffering from the atom loss process with a rate $\gamma_{los}= 5$ Hz. Panel (b) shows the repeated result for the system, where the atom loss is compensated by the atom injection with a rate $\kappa_{pump}=400$ kHz, resulting to a steady-state number of atoms $N_s =\kappa_{inj}/\gamma_{loss}=8\times10^4$. } 
\end{figure}

In the main text, we pointed out that the frequency uncertainty is currently limited by the atom loss process, which determines the time for single measurement. To overcome this problem, we propose to compensate the loss by injecting new atoms. In \citep{YZhang-1}, we have modeled the atom loss and injection as stochastic jumps between the Dicke states with varying number of atoms.  Here, we model within the second-order mean-field theory as detailed in the main text with an ancillary rate-equation for the number of atoms  $\partial N_i /\partial t = - \gamma_{los,i} N_i + \lambda_{inj,i} $, where  $ \gamma_{los,i},\lambda_{inj,i} $ describe the atom loss rate and injection injection rate, respectively.

In Fig. \ref{fig:atomlosspumping}, we investigate the system suffering from the atom loss (a) and the system (b) with the atom loss compensated by the atom injection. Fig. \ref{fig:atomlosspumping} (a) shows a superradiance pulse with a maximum of $1.2$ photons at around $90$ ms. The inset shows that the total number of atoms decays exponentially from the initial value of $9\times 10^4$ to about  $2\times 10^4$ at $300$ ms and the number of excited (un-excited) atoms follows the trend of the total number for the earlier (later) time. Because of the reduced number of atoms, the superradiance is about 2 times smaller than the system in the absence of atom loss (not shown). Here, we apply strong laser pulses and excite the cavity in a very short time which causes the rapid dynamics of the photon number and the ground and excited state atomic populations. The figure also shows the influence of atom loss in single experiment. The most serious obstacle for the experiment is the need to reload the atoms which takes much longer than the duration of the measurement of the superradiance pulses. 

Fig. \ref{fig:atomlosspumping} (b) shows that for the case of continuous atom injection the number of atoms converges to a steady-state value $8\times10^4$ for longer time (see the inset), and the superradiance pulses can be repeated several times. As a result, we can directly initialize the atomic ensemble after each superradiant emission stage, which reduces dramatically the time per measurement to about $0.25$ s. In view of Fig. \ref{fig:beats-two-ensembles}, we except the fractional Allan deviation $7\times 10^{-17}/\sqrt{\tau/s}$ for the frequency measurement with the system studied here.  

\begin{figure}[t]
\begin{centering}
\includegraphics[scale=0.28]{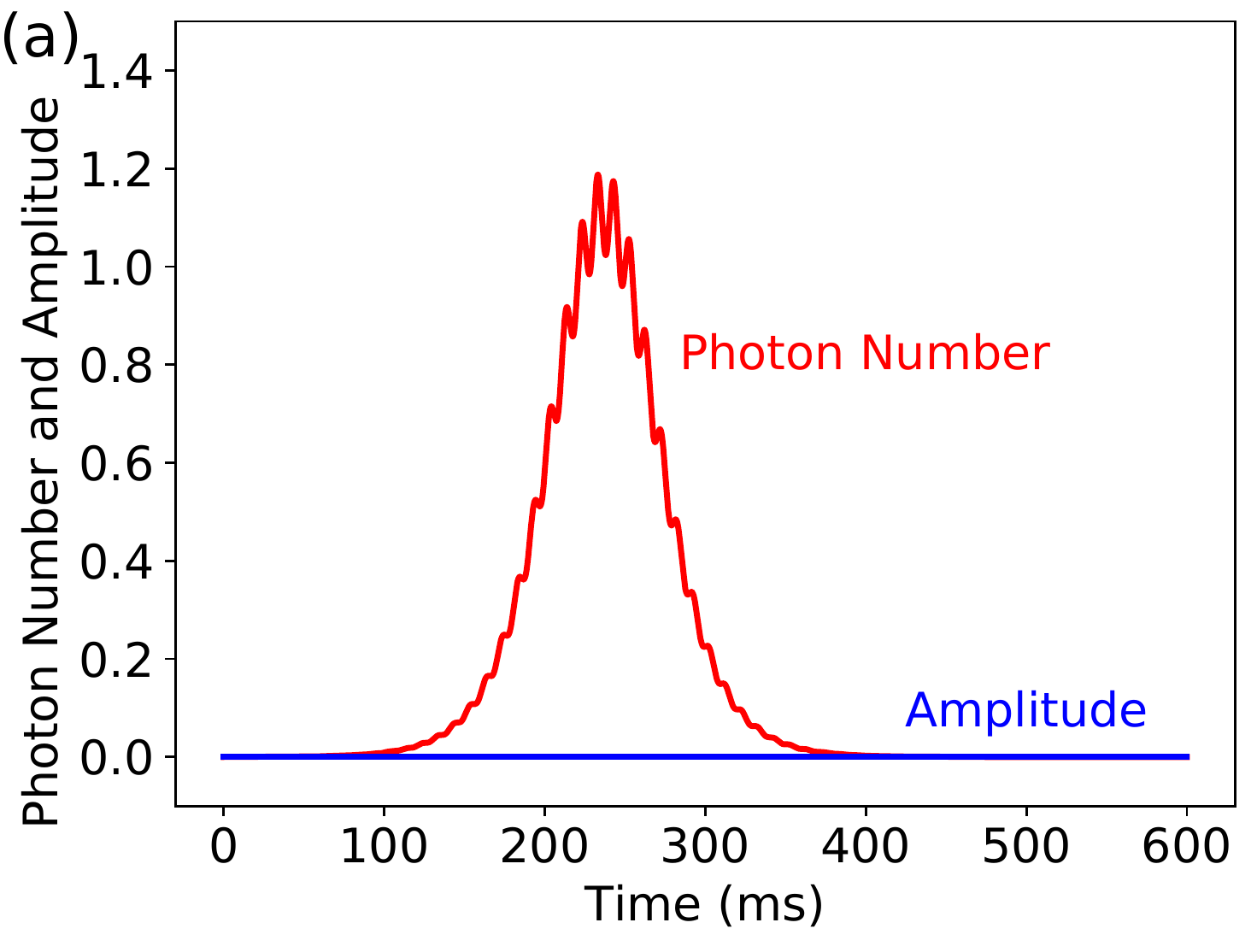}
\includegraphics[scale=0.28]{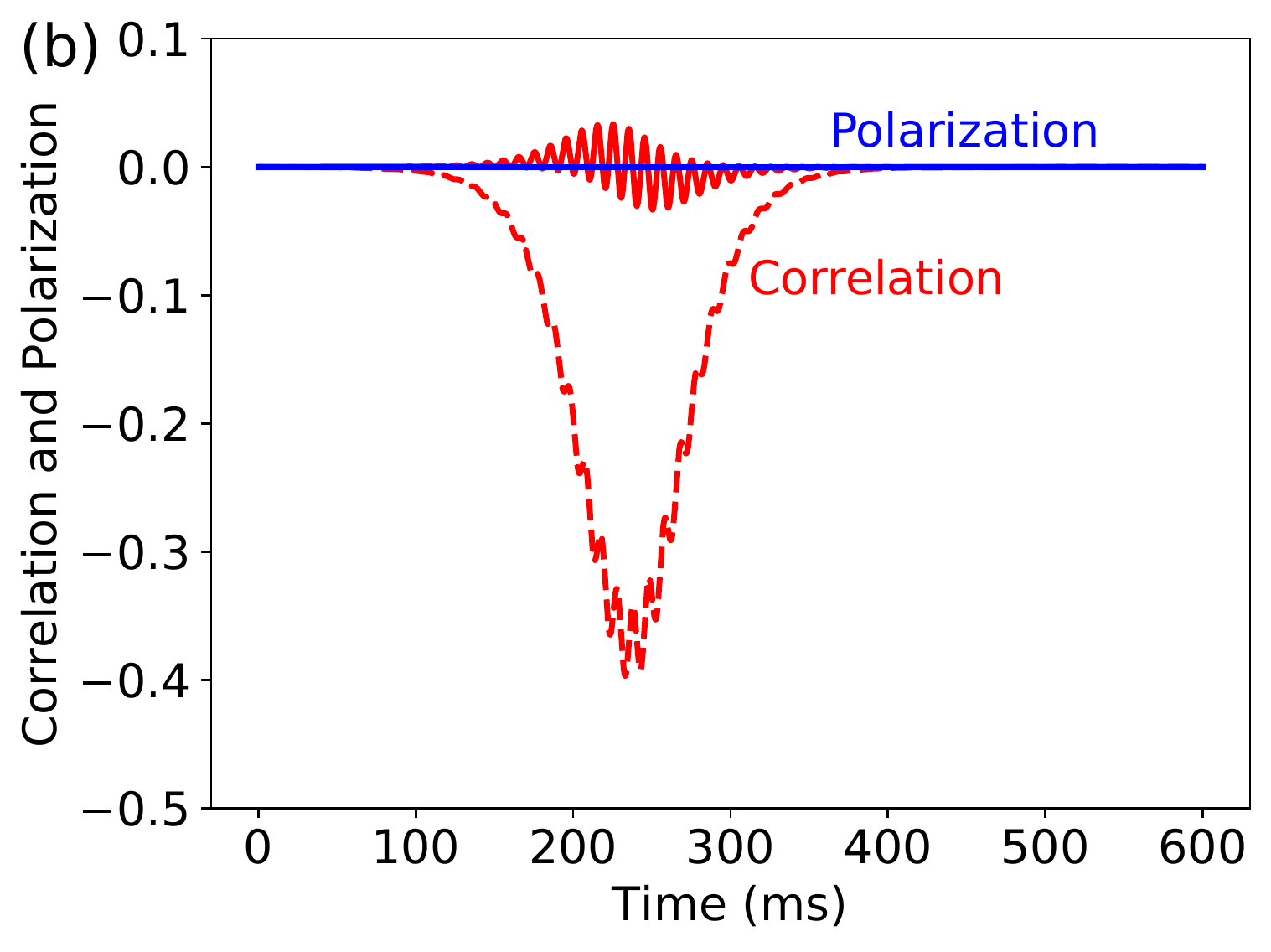}
\par\end{centering}
\begin{centering}
\includegraphics[scale=0.28]{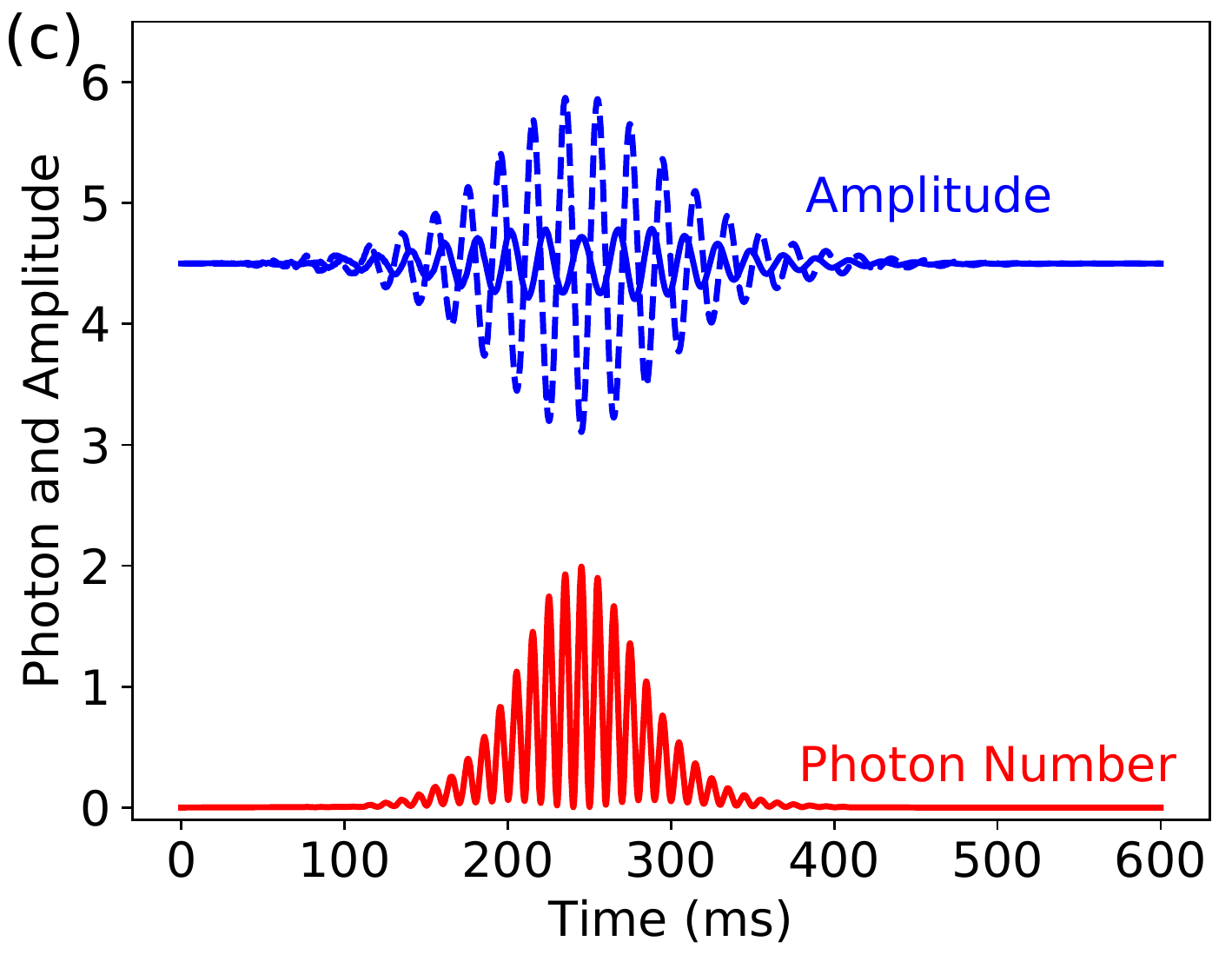}
\includegraphics[scale=0.28]{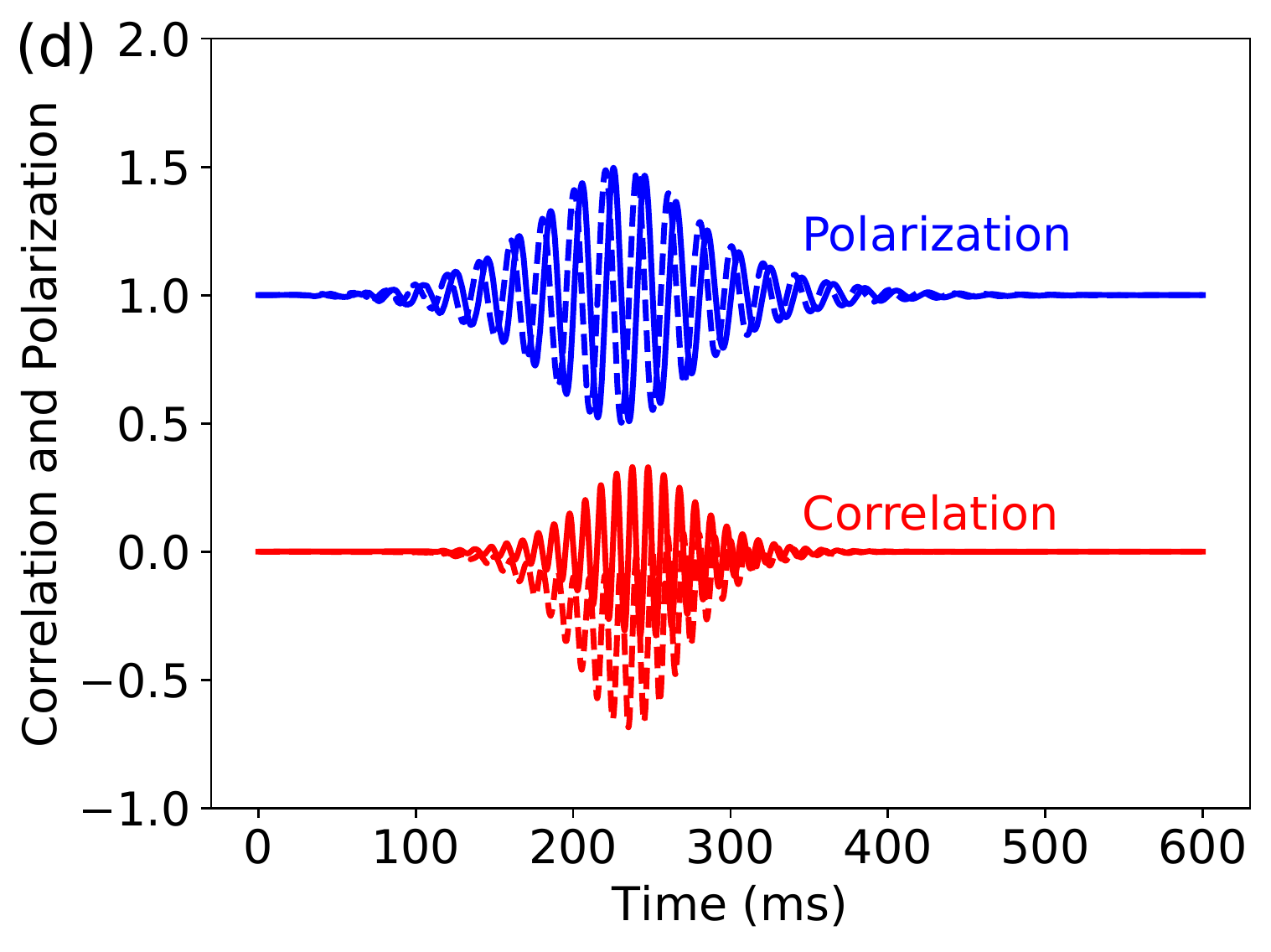}
\par\end{centering}
\caption{\label{fig:particular-case} The mean values of field and atomic observables during superradiant decay from the fully excited state. The physical parameters are the same as in Fig. \ref{fig:optimizedradiation}.
Panels (a,b) show the vanishing mean value of the optical amplitude $\langle a\rangle$ and atomic polarization $\left \langle \hat{\sigma}_i^- \right \rangle$ (blue lines) and the photon number and atom-photon  correlations $\left \langle a \hat{\sigma}_i^+ \right \rangle$ (red lines) in the absence of probing. Panels (c,d) show the same quantities (solid: real part, dashed: imaginary part), conditioned on heterodyne detection. In panels (c,d), we shift the blue lines vertically for the sake of clarity.}
\end{figure}

\subsection{Comparison of Dynamics with and without Heterodyne Detection \label{subsec:comparison}}

Our numerical simulations of the quantum dynamics yield realistic heterodyne detection records, which are then subject to frequency analysis as applied in experiments. While the instantaneous measurement noise is modelled by a Wiener process in Eq.\eqref{eq:measurment-backaction}, the measurement back action causes transients and correlations with the signal values at later times. It is interesting to note that in the simulations we obtain non-vanishing mean fields due to the measurement back action, while the unconditional  second order mean value equations account for the average dynamics of the atoms and fields without breaking the phase symmetry and without introducing first order mean fields. Similarly, the average value of temporal correlations of detected signals, which are obtained in processes that break the symmetry, is reproduced by the deterministic master equation and the quantum regression theorem which assume only non-vanishing  second order mean values, see \cite{qrtnoise}. 

To illustrate this point, in Fig. \ref{fig:particular-case} (a,c), we compare the intra-cavity photon number (red solid lines) and the cavity field amplitude (blue lines) in the absence and presence of heterodyne detection, respectively. When averaged over many simulated experiments, the results of the stochastic master equation should follow the unconditional density matrix dynamics. This equivalence, however, is only exact for the the linear Lindblad master equation, and the replacement of third-order operator products with first and second order terms in the mean field approximation might introduce deviations from this equivalence. Still, Fig. \ref{fig:particular-case} shows that the intra-cavity photon number and the atom-photon correlation have comparable pulse structures in the unconditional and the conditioned dynamics, while the stochastic realization of the cavity field amplitude differs appreciably.

\begin{figure}[t]
\begin{centering}
\includegraphics[scale=0.45]{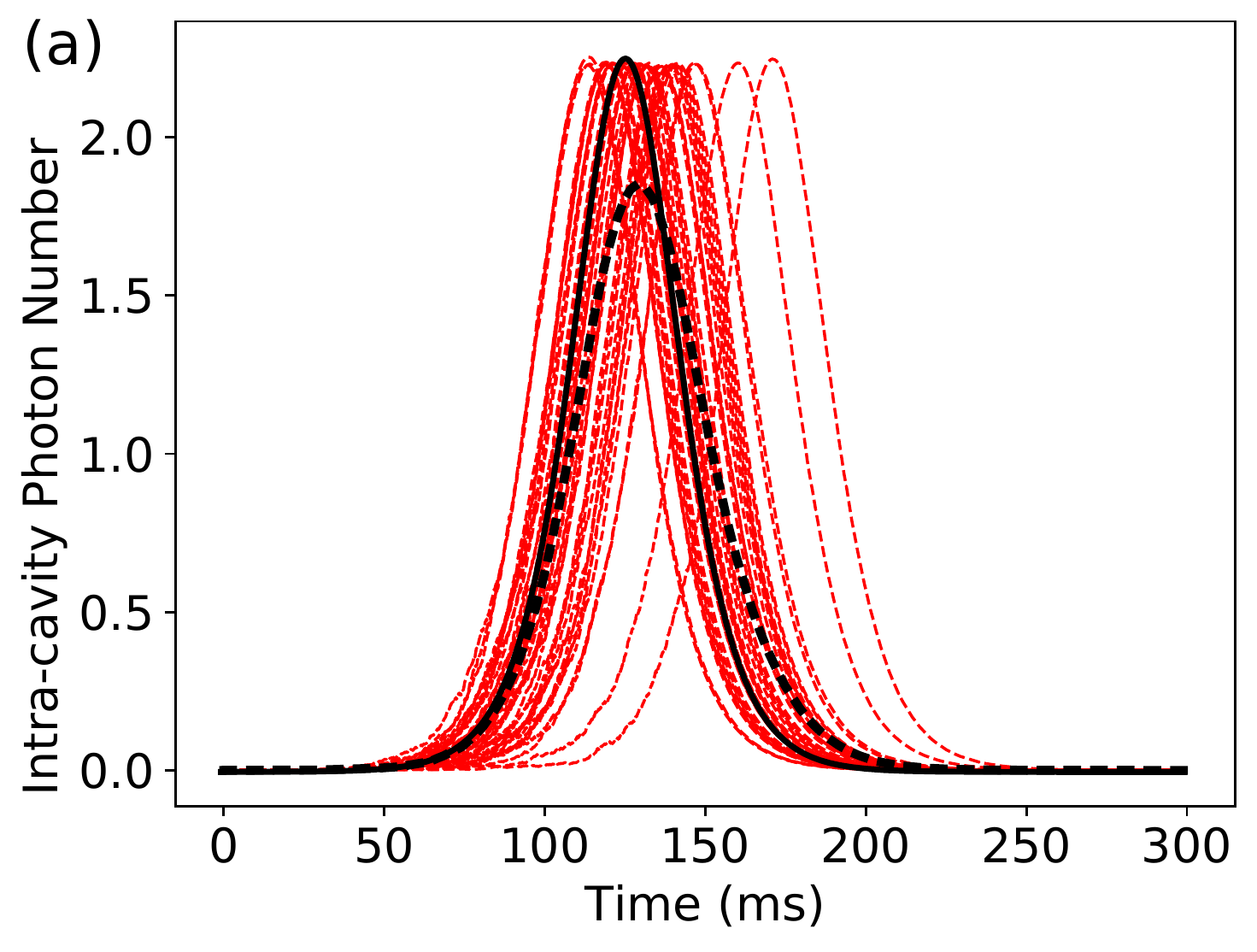}
\includegraphics[scale=0.45]{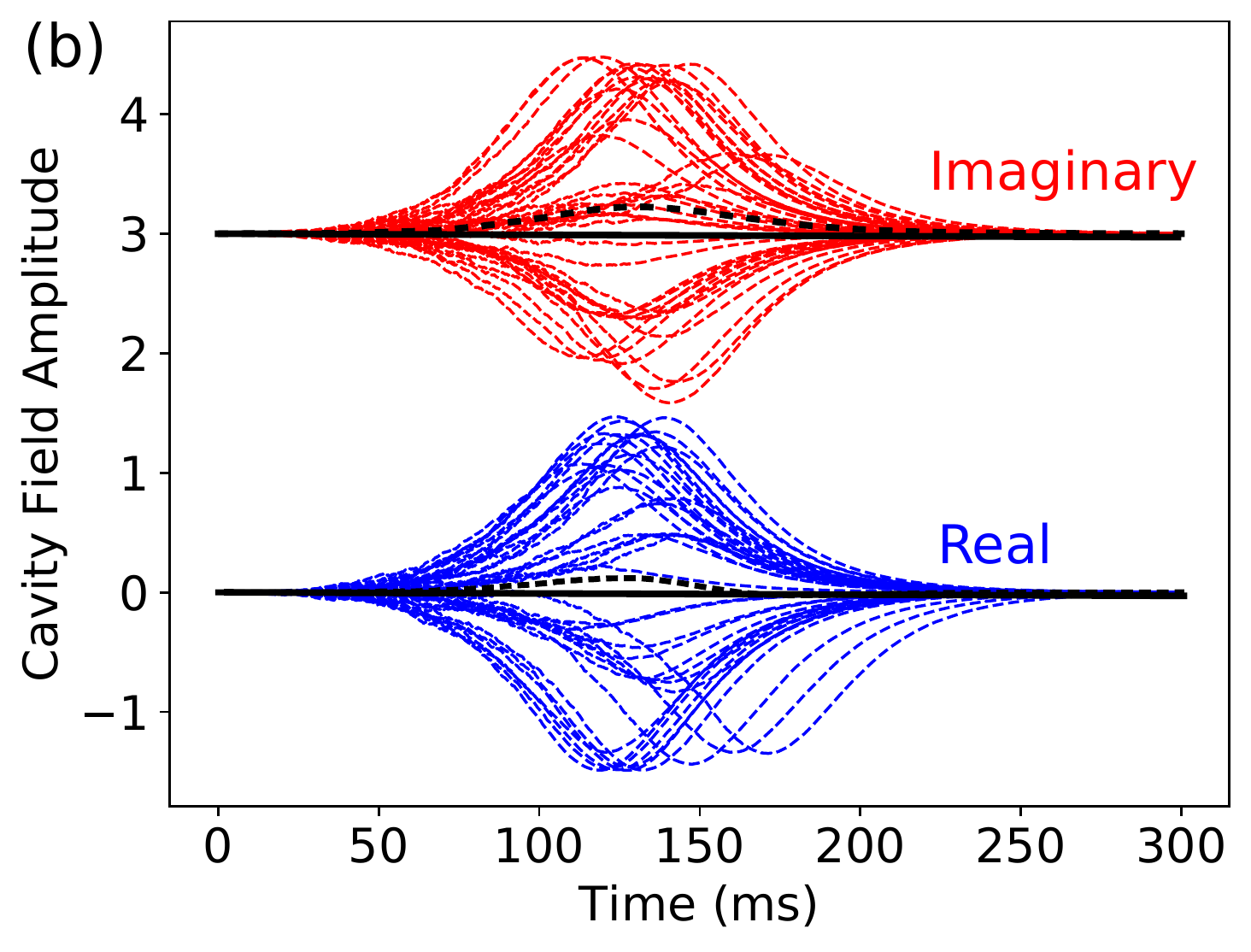}
\par\end{centering}
\caption{\label{fig:conditionaldynamics}  Comparison of the dynamics of the monitored and un-monitored system as in Fig. \ref{fig:particular-case} with a vanishing magnetic field and the other parameters as listed in Fig. \ref{fig:optimizedradiation}.  Panel (a) shows 40 trajectories of the intra-cavity photon number (red thin dotted line), their average (blue dashed line), and the trajectory for an un-monitored system (blue solid line). Panel (b) shows the real (blue lines) and imaginary part (red lines, shifted vertically for the sake of clarity) of the cavity field amplitude. }
\end{figure}

As a further illustration, in Fig. \ref{fig:conditionaldynamics} we show results of  40 independent simulations and show that the simulated intra-cavity field amplitude develops different complex phases and their average is close to  zero, as determined for the un-monitored system.  The deviation of the average results from the un-monitored dynamics may be both due to the limited number of trajectories and the handling of non-linear terms in the mean field theory. We recall, that the frequency analysis is not based on the time dependent mean values shown in these figures but on a Fourier analysis of the simulated heterodyne signal.

\end{document}